\documentclass[%
twocolumn,
superscriptaddress,
nofootinbib,
 amsmath,amssymb,
 aps, prd
]{revtex4-2}

\usepackage{graphicx}
\usepackage{tensor}
\usepackage{siunitx}
\usepackage{xcolor} 
\usepackage[normalem]{ulem} 
\usepackage{lipsum} 

\usepackage[pdfborder={0 0 0}, plainpages=false, hyperfigures=true]{hyperref}

\newcommand{\beq}{\begin{equation}}
\newcommand{\eeq}{\end{equation}}
\newcommand{\beqn}{\begin{eqnarray}}
\newcommand{\eeqn}{\end{eqnarray}}

\newcommand{\lo}{\mathrel{\raise.3ex\hbox{$<$}\mkern-14mu
    \lower0.6ex\hbox{$\sim$}}}
\newcommand{\go}{\mathrel{\raise.3ex\hbox{$>$}\mkern-14mu
    \lower0.6ex\hbox{$\sim$}}}

\newcommand{\WSU}{\affiliation{Department of Physics \& Astronomy,
	Washington State University, Pullman, Washington 99164, USA}}
\newcommand{\UNH}{\affiliation{Department of Physics \& Astronomy, University of New Hampshire, 9 Library Way, Durham NH 03824, USA}}
\newcommand{\TAPIR}{\affiliation{TAPIR, Walter Burke Institute for Theoretical Physics, MC 350-17, California Institute of Technology, Pasadena, California 91125, USA}}
\newcommand{\Cornell}{\affiliation{Cornell Center for Astrophysics and Planetary Science, Cornell University, Ithaca, New York, 14853, USA}}
\newcommand{\MPI}{\affiliation{Max Planck Institute for Gravitational Physics (Albert Einstein Institute), D-14467 Potsdam, Germany}}

\newcommand{\NCSA}{\affiliation{National Center for Supercomputing Applications, University of Illinois, 1205 W Clark St, Urbana IL 61801, USA}}
\newcommand{\UIUCPhys}{\affiliation{Department of Physics, University of Illinois, 1110 West Green St, Urbana IL 61801, USA}}

\newcommand{\Barnard}{\affiliation{Department of Physics and Astronomy, Barnard College, 3009 Broadway, Altschul Hall 504A, New York, NY 10027}}

\begin{document}
\title{General relativistic simulations of collapsing binary neutron star mergers with Monte-Carlo neutrino transport}
\author{Francois Foucart}\UNH
\author{Matthew D. Duez}\WSU
\author{Roland Haas}\NCSA\UIUCPhys
\author{Lawrence E. Kidder}\Cornell
\author{Harald P. Pfeiffer}\MPI
\author{Mark A. Scheel}\TAPIR
\author{Elizabeth Spira-Savett}\UNH\Barnard

\begin{abstract}

Recent gravitational wave observations of neutron star-neutron star and neutron star-black hole binaries appear to indicate that massive neutron stars may not be too uncommon in merging systems. These discoveries have led to an increased interest in the simulation of merging compact binaries involving massive stars. In this manuscript, we present a first set of evolution of massive neutron star binaries using Monte-Carlo radiation transport for the evolution of neutrinos. We study a range of systems, from nearly symmetric binaries that collapse to a black hole before forming a disk or ejecting material, to more asymmetric binaries in which tidal disruption of the lower mass star leads to the production of more interesting post-merger remnants. For the latter type of systems, we additionally study the impact of viscosity on the properties of the outflows, and compare our results to two recent simulations of identical binaries performed with the WhiskyTHC code. We find agreement on the black hole properties, disk mass, and mass and velocity of the outflows within expected numerical uncertainties, and some minor but noticeable differences in the evolution of the electron fraction when using a subgrid viscosity model, with viscosity playing a more minor role in our simulations. The method used to account for r-process heating in the determination of the outflow properties appears to have a larger impact on our result than those differences between numerical codes. 
We also use the simulation with the most ejected material to verify that our newly implemented Lagrangian tracers provide a reasonable sampling of the matter outflows as they leave the computational grid. We note that, given the lack of production of hot outflows in these mergers, the main role of neutrinos in these systems is to set the composition of the post-merger remnant. One of the main potential use of our simulations is thus as improved initial conditions for longer evolutions of such remnants.
 
\end{abstract}

\maketitle

\section{Introduction}

The last five years have offered a sudden wealth of information about the properties of merging compact objects, thanks to gravitational wave observations by the LIGO-Virgo-KAGRA collaboration. The latest catalogues of gravitational wave events
~\cite{LIGOScientific:2018mvr,LIGOScientific:2020ibl,LIGOScientific:2021usb,LIGOScientific:2021djp} include in particular two binary neutron star mergers and $2-5$ black hole-neutron star mergers. Interestingly, the neutron stars observed in these systems include objects of higher mass than what one might have expected from the observed population of neutron stars in galactic binaries. Galactic observations of compact neutron star binaries favor low-mass systems~\cite{Ozel2012}, even though higher mass neutron stars are observed in e.g. the population of millisecond pulsars~\cite{Antoniadis:2016}. In gravitational wave observations, we now have a $\sim 3.4M_\odot$ binary neutron star system (GW190425), a $(1.7-2.2)M_\odot$ neutron star in a neutron star-black hole binary (GW200105 162426), and lower mass objects in the neutron star binary GW170817 and the neutron star-black hole binary GW191219 163120. Other observed neutron stars have poorly measured masses, and a few observed compact objects may be either high mass neutron stars or low mass black holes (e.g. GW200210 092254). Overall, we thus observe a much wider range of masses than initially expected.

As a result, while early simulations of neutron star binaries focused mostly on low-mass systems, there has been an increased interest in recent years in modeling high-mass binaries (see e.g.~\cite{Bauswein:2020aag,Bernuzzi:2020txg,Dudi:2021abi,Camilletti:2022jms}), especially considering existing deficiencies in the semi-analytical models used to predict the outcome of neutron star mergers in that less studied part of the available parameter space~\cite{Camilletti:2022jms,Henkel:2022naw}. Parameter space coverage for these high mass systems remains however sparser than for lower mass systems.

In this manuscript, we focus specifically on systems that collapse to a black hole within $\sim 1\,{\rm ms}$ of the collision between the two neutron star cores. There are effectively two known regimes for such systems. For near equal-mass systems, there remains only a small amount of material outside of the forming black hole with enough angular momentum to avoid being immediately accreted by the black hole remnant. These systems typically result in nearly no mass ejection, and low-mass accretion disks. For more asymmetric systems, on the other hand, the lower mass neutron star may be of sufficiently low compaction to be partially disrupted before merger. In that case, material in the tidal tail formed during merger has enough angular momentum to avoid falling into the black hole. Mass ejection and the formation of massive disks are then possible.

All of our simulations are performed with the SpEC code~\cite{SpECwebsite}, using a self-consistent evolution of Einstein's equations and of the equations of general relativistic hydrodynamics~\cite{Duez:2008rb}. We also use our new state-of-the-art Monte-Carlo radiation transport code~\cite{Foucart:2020qjb} to account for neutrino-matter interactions and neutrino transport. We note that Monte-Carlo transport is particularly convenient to use for the post-merger evolution of these systems, as Monte-Carlo methods are very efficient in the absence of a dense neutron star in the system and our remnants are all black hole-disk systems. We simulate a near equal-mass system, as well as a range of unequal mass binaries, and focus particularly on the properties of the remnant black hole, the accretion disk, and the matter and neutrino outflows. Two of our systems are specifically chosen to match the binary parameters used in~\cite{Bernuzzi:2020txg}, allowing for more detailed comparison of simulations with very different numerical methods. We will see that our results generally show agreement between the two codes within expected numerical errors, with the possible exception of the electron fraction of some outflows. Differences in the electron fraction may be due to different viscosity models or neutrino transport algorithms in the simulations.

For the simulation producing the most massive disk and outflows, we also perform a more detailed study of the impact of subgrid viscosity on the merger, using our recent implementation of a LES (Large Eddy Simulation) model~\cite{Duez:2020lgq}. That model is a modification of the viscosity model of Radice {\it al}~\cite{Radice:2017} used in~\cite{Bernuzzi:2020txg}, with a few key differences detailed in the methods section. We also use this simulation to perform a detailed analysis of a new implementation of Lagrangian tracers in SpEC, aimed at determining whether differences between the evolution of the tracers and the evolution of the fluid lead to the tracers not being a good sampling of the matter outflows at the end of the simulation. This new implementation is designed to improve load-balancing of the tracer evolution, allowing us to follow a larger number of Lagrangian particles, but has the disadvantage of using a lower-order time-stepping algorithm than the fluid evolution. For the neutron star merger considered here, we find that the tracers remain a reasonably good sampling of the fluid (i.e. with errors of the same order as expected sampling errors), yet we caution that we found much less encouraging results in simulations of a high-mass black hole-neutron star system. 

Finally, we note that the previously mentioned viscosity model is meant to capture the effective impact of the Kelvin-Helmholtz instability during merger. After merger, the main source of effective viscosity in the accretion disk is expected to be the magnetorotational instability (MRI). Accordingly, for the two simulations forming the most massive disks, we switch to a subgrid model for viscosity calibrated to the MRI $5\,{\rm ms}$ after merger (following~\cite{Haddadi:2022qcu}), continuing these simulations up to $10\,{\rm ms}$ post-merger in order to reach a relatively settled black hole-disk post-merger remnant that can be used as initial condition for longer post-merger simulations.

\section{Overview of the simulations}
\label{sec:simulations}

In this manuscript, we consider a total of four distinct binary systems, described in more detail in this section.
First, we simulate three binaries in which the equation of state of dense matter is modeled using the SFHo equation of state~\cite{Steiner:2012rk}. The neutron stars in these binaries have gravitational masses ($m_1,m_2$) = ($1.61M_\odot$,$1.51M_\odot$), ($1.80M_\odot$,$1.31M_\odot$), and ($1.78M_\odot$,$1.06M_\odot$). We then repeat the last configuration with the LS220 equation of state~\cite{Lattimer:1991nc}.

\begin{table*}
\begin{tabular}{c|cccccccc}
Name & $M_1$ & $M_2$ & EoS  & $R_1$ (km) & $R_2$ (km) & $\tilde \Lambda$ \\
\hline
SFHo-161-151 & $1.61M_\odot$ & $1.51M_\odot$ & SFHo & 11.7 & 11.8 & 170\\
SFHo-180-131 & $1.80M_\odot$ & $1.31M_\odot$ & SFHo & 11.5 & 11.9 & 360\\
SFHo-178-106 & $1.78M_\odot$ & $1.06M_\odot$ & SFHo & 11.5 & 12.0 & 1450\\
\hline
LS220-178-106 & $1.78M_\odot$ & $1.06M_\odot$ & LS220 & 12.2 & 12.8 & 2460\\
\end{tabular}
\caption{Summary of the configurations simulated in this manuscript. We list the name of the simulation, the gravitational mass of each neutron star, the equation of state, the radius of each star, and the reduced tidal deformability of the binary. The last configuration was simulated at two resolutions, as well as with viscosity turned on/off at different times in the evolution.}
\label{tab:id}
\end{table*}

These systems represent three potential outcome of binary neutron star mergers with rapidly collapsing remnants. In the first case, nearly all of the matter rapidly falls into the forming black hole, leaving a disk of negligible mass around it and nearly no mass ejection. In the second case, a low-mass disk ($0.04M_\odot$) forms around the remnant black hole, and we still do not observe significant mass ejection during merger. In the third and fourth cases, tidal disruption of the low-mass star by its massive companion leads to the formation of a more massive disk ($\sim 0.15M_\odot$) and significant mass ejection in a cold tidal tail ($\gtrsim 0.001M_\odot$). We note that while the least massive neutron star in these two configurations has a very low mass, these systems were chosen because they match simulations performed in~\cite{Bernuzzi:2020txg} with a different code and different treatment of neutrinos and viscosity. This allows for easy comparisons with our results, and additionally provides us with an extreme case that should be useful in the construction of analytical models for remnant disks and matter outflows. To test the robustness of our results, we perform the LS220 simulation at two different resolutions, and with or without subgrid viscosity. Specifically, we ran simulations without viscosity; with viscosity included only after black hole formation; with viscosity included only up to the contact; and with viscosity included during the entire simulation. The second and third case here are meant to verify that the subgrid viscosity does not have an important effect at times when the Kelvin Helmotz instability is not active. The binary using the LS220 equation of state is also used to test our implementation of tracer particles. Finally, we studied the impact of the number of Monte-Carlo packets during the inspiral, collapse to a black hole, and post merger evolution. We note however that most of those changes had only minor impacts on the outcome of the simulations for the observables discussed in this manuscript, with the exception of the number of packets used during the collapse itself (see below).

Initial configurations for our simulations are constructed using the Spells code~\cite{Pfeiffer2003,FoucartEtAl:2008}. We initially build quasi-circular configuration (i.e. stars without radial velocity at the initial time), which results in residual eccentricities $e\sim 0.01$, then perform one round of eccentricity reduction to obtain orbits with $e \lesssim 0.003$, following the method described in~\cite{Pfeiffer-Brown-etal:2007}.  The near-equal mass system is initialized $\sim 4$ orbits before contact, while the other three binaries are evolved for $6-7$ orbits before contact. We evolve the post-merger remnants for $4\,{\rm ms}$ after collapse to a black hole, long enough to be able to extract values for the mass of the post-merger disk and of the dynamical ejecta (if any ejecta is produced), as discussed in Sec.~\ref{sec:results}. In addition, the two most asymmetric configurations are followed up to $10\,{\rm ms}$ post-collapse with a subgrid viscous model meant to approximately capture heating and angular momentum transport due to the MRI.

In the rest of this manuscript, we refer to the four binary systems described in this section using the names SFHo-161-151, SFHo-180-131, SFHo-178-106, and LS220-178-106, i.e. the equation of state followed by the mass of each star, in units of $0.01M_\odot$. A list of all  physical configurations is provided in Table~\ref{tab:id}. 

\section{Numerical methods}
\label{sec:methods}

\subsection{General Relativistic Hydrodynamics}

All simulations are performed using the Spectral Einstein Code (SpEC)~\cite{SpECwebsite}. SpEC evolves the spacetime metric using the Generalized Harmonics 
formulation of Einstein's equations using pseudospetral methods~\cite{Lindblom:2007},
and the general relativistic equations of hydrodynamics on a separate finite volume grid~\cite{Duez:2008rb}. A number of different evolution algorithms are available in SpEC,
to accommodate both simulations with a low-level of microphysics requiring less dissipative, high-order methods in order to decrease numerical errors (for waveform modeling), and simulations with more detailed microphysics for which phase accuracy during the evolution is not as crucial and slightly more dissipative methods are preferable for stability. The simulations presented here fall in the second category.

The fluid equations are evolved in conservative form
using high-order shock capturing methods (HLL approximate Riemann solver~\cite{HLL} and WENO5 reconstruction from cell centers to cell faces~\cite{Liu1994200,Jiang1996202}).
Both systems of equations are evolved in time using third-order Runge-Kutta time stepping. The evolution of the metric uses adaptive step sizes, while the evolution of the fluid uses time steps $\Delta t = 0.25 \min{(\Delta x/c_g)}$, with $c_g$ the speed of light in grid coordinates, and $\Delta x$ the grid spacing. The minimum is taken
over all cells in our computational domain. Given the stricter time stepping constraints on the spectral grid, this results in larger desired time steps on the finite
volume grid than on the pseudospectral grid. In practice, we require an integer number of time steps on the pseudospectral grid for every step on the finite volume grid. At the end of each time step on the finite volume grid, we communicate metric quantities from the pseudospectral grid to the finite volume grid, and fluid variables in the reverse direction. Values of these quantities at other times are, when needed, extrapolated in time from the last two communicated values.

The Generalized Harmonics formulation of Einstein's equation requires the prescription of a source term $H^a$ for the evolution of the coordinates $x^b$, which follows the inhomogeneous wave equation
\beq
g^{ab} \nabla^c \nabla_c x^b = H_a(\bar x, g_{ab}),
\eeq
with $g_{ab}$ the spacetime metric. The initial value of $H_a$ is set by requiring that $g^{tt}$ and $g^{ti}$ are constant in the coordinate system corotating with the binary at the initial time. We then smoothly transition first to the harmonic condition $H_a=0$ during inspiral, and then to the ``damped harmonic'' prescription from Szilagyi et al.~\cite{Szilagyi:2009qz} after contact. Once an apparent horizon is found in our simulation, we additionally excise the region inside of that horizon, as described in~\cite{Haas:2016}. More detail on our numerical methods can be found in~\cite{Duez:2008rb,Foucart:2013a}.

\subsection{Equation of state}
\label{sec:eos}

The fluid equations require us to prescribe an equation of state for the dense matter within the neutron stars. We use tabulated versions of the SFHo~\cite{Steiner:2012rk} and LS220~\cite{Lattimer:1991nc} equations of state, taken from the StellarCollapse library~\cite{OConnor2010}. These tables provide us with the internal energy, pressure, and sound speed of the fluid as a function of baryon density $\rho_0$, temperature $T$, and electron fraction $Y_e$. The LS220 equation of state leads to $1.4M_\odot$ neutron stars with a radius of $12.7\,{\rm km}$, and a maximum mass for non-rotating neutron stars in isolation of $2.04M_\odot$. The SFHo equation of state leads to more compact stars ($11.9\,{\rm km}$ for $1.4M_\odot$ stars) and a similar maximum mass ($2.06M_\odot$). Both lead to macroscopic properties for neutron stars that are consistent with current astrophysical constraints, even though they are not compatible with all nuclear physics constraints at lower density. We note that the definition of the baryon density is not entirely consistent in different equations of state tables: we define $\rho_0 = m_b n_b$ with $n_b$ the number density of baryons and $m_b$ the assumed mass of a baryon. Numerical simulations use a constant $m_b$ in evolutions because they evolve the baryon density of the fluid, but the true conserved quantity is the number of baryons; that is we evolve the equation
\beq
\nabla_a (m_b n_b u^a) = 0
\eeq
with $u^a$ the 4-velocity of the fluid, taking advantage of the fact that at constant $m_b$, this is just the equation for the conservation of baryon number. Any reasonable choice for $m_b$ (proton mass, neutron mass, average mass of a nucleon in a given nucleus) is however acceptable. This does not impact the evolution of the fluid, which only cares about $n_b$ and the total energy density $u=\rho_0(1+\epsilon)$, with $\epsilon$ the specific internal energy of the fluid -- but it does impact what our simulation considers to be "rest mass energy" ($\rho_0$) vs "internal energy" ($\rho_0 \epsilon$). This distinction becomes relevant when attempting to determine the fate of the ejecta, as discussed in Sec.~\ref{sec:ejecta}. In our tables, the reference density for the SFHo equation of state is close to the average mass per baryons of the nuclei formed in r-process nucleosynthesis, while for the LS220 equation of state it is the neutron mass.

\subsection{Subgrid viscosity}

During the merger of two neutron stars and the post-merger evolution of neutron star-disk or black hole-disk systems, we expect significant growth of (magneto)hydrodynamical instabilities. The Kelvin-Helmholtz instability is active at the interface between the two colliding stars during merger. After merger, the MRI is active in the remant disk as well as, possibly, in some regions of the remnant neutron star. These instabilities drive small scale turbulence that is expected to play an important role for angular momentum transport and heating of the post-merger remnant. If a dynamo mechanism is active in the remnant, they may also eventually lead to the production of strong magnetically-driven disk winds and relativistic jets~\cite{Rezzolla:2011da,ruiz:16,Siegel:2017nub,Fernandez:2018kax,Mosta:2020hlh}. Capturing the growth of these small scale instabilities is however very costly, and getting converged predictions for the large scale magnetic field post-merger is beyond even the highest resolution simulations performed so far~\cite{Kiuchi2015}. A possible alternative to at least qualitatively capture angular momentum transport and heating in the remnant is to rely on approximate subgrid models explictly introducing viscosity in the simulations. A few classes of subgrid models have been used in merger simulations so far: a relativistic $\alpha$-viscosity model~\cite{Fujibayashi:2017puw} adapted from the Israel-Stewart formalism for non-ideal fluids~\cite{ISRAEL1979341}, a simpler (but not fully covariant) turbulent mean-stress model~\cite{Radice:2017}, and the gradient sub-grid scale model of~\cite{Palenzuela:2021gdo}. In the first two cases, the stress-energy tensor of the ideal fluid
\beq
T_{ab,{\rm ideal}} = \left[\rho_0 (1+\epsilon)+P\right] u_a u_b + P g_{ab}
\eeq
(with $P$ the pressure of the fluid) is complemented by an approximate non-ideal piece $\tau_{ab}$:
\beq
T_{ab} = T_{ab,{\rm ideal}} + \tau_{ab}.
\eeq
In~\cite{Palenzuela:2021gdo}, the magnetic field is directly evolved and a subgrid magnetic stress tensor is added to the evolution equations instead.

In this work, some of our simulations use the shear viscosity model of Radice~\cite{Radice:2017}, modified to match the expected Newtonian limit for the energy equation and to impose the expected condition $\tau_{ab}u^a=0$~\cite{Duez:2020lgq}. The spatial components of $\tau_{ab}$ are taken from~\cite{Radice:2017}:
\beq
\tau_{ij} = -2 \rho_0 l c_s h W^2\left[ \frac{1}{2} \left(\nabla_i v_j + \nabla_j v_i\right)-\frac{1}{3} \nabla_k v^k \gamma_{ij}\right]
\eeq
with $h=1+\epsilon+P/\rho_0$ the specific enthalpy, $c_s$ the sound speed, $W$ the Lorentz factor of the fluid, $v^i$ its 3-velocity, and $\gamma_{ij}$ the 3-metric on a slice of constant time coordinate. Given a normal vector $n^a$ to that slice, the 3-velocity is given by $u^a = W(n^a + v^a)$, and the 3-metric by $\gamma_{ab}=g_{ab}+n_a n_b$. We can see from this definition that the method is not fully covariant, as it requires the choice of a preferred time direction, and uses covariant derivatives of the velocity on the spatial slice orthogonal to that preferred direction.

The length scale $l$ sets the strength of non-ideal effects, and should be calibrated to the result of simulations capturing magnetic turbulence in the post-merger remnant. In our simulations using viscosity, during merger, we consider the simple choice $l=30\,{\rm m}$, which is expected to be at the upper end of the range of reasonable values for the Kelvin-Helmholtz instability in neutron star mergers~\cite{Radice:2017}. We note however that more advanced models in which $l$ depends explicitly on the fluid density have been developed in~\cite{radice2020binary}, and a density-dependent mixing length is in particular using in the simulations of~\cite{Bernuzzi:2020txg} that we are here using as a comparison point. While our two viscosity schemes are thus fairly similar, they differ both due to our correction to the time components of the stress-tensor and due to the choice of a smaller mixing length at low-density in~\cite{Bernuzzi:2020txg}. As our simulations find a lesser impact of viscosity on the properties of the outflows, and we use a larger mixing length, it is however unlikely that the choice of mixing length alone explains the small differences between the two sets of simulations.

For simulations continued up to $10\,{\rm ms}$ post merger, we modify the value of $l$ to more closely match the expected effect of the MRI. Following~\cite{Haddadi:2022qcu}, we choose
\beq
l = \alpha_{\rm vis} \frac{P}{\Omega_K c_S \rho_0}
\eeq
with $\alpha_{\rm vis}=0.03$. We switch to this viscosity model $5\,{\rm ms}$ post-merger, after formation of a clear accretion disk.

This subgrid model is expected to qualitatively capture heating and angular momentum transport. It does not however capture the effects of a large scale magnetic field. Accordingly, we cannot use this model to study relativistic jets, gamma-ray bursts, or magnetically-driven winds. High-resolution 3D MHD simulations would be required to accomplish those objectives.

Unless otherwise noted, all figures in this manuscript for simulations LS220-178-106 and SFHo-178106 are for the simulations including a subgrid viscosity model. The simulations without viscosity performed for case LS220-178-106 are only used to estimate the impact of that model.

\subsection{Neutrino transport}

Neutrinos are evolved using our recently developed Monte-Carlo algorithm~\cite{Foucart:2021mcb}. In this algorithm, neutrino packets directly sample the distribution function of neutrinos, allowing for the evolution of Boltzmann's equations of radiation transport. Packets are emitted isotropically in the fluid frame, sampling the emission spectrum of neutrinos, then propagate along geodesics. During this propagation, each packet has a finite probability of being absorbed or scattered by the fluid, determined by tabulated values of the absorption and scattering opacities. We consider 3 species of neutrinos: electron neutrinos $\nu_e$, electron antineutrinos $\bar\nu_e$, and heavy-lepton neutrinos $\nu_x$ (which include muon and tau neutrinos and antineutrinos). We use tables generated by the public code NuLib~\cite{OConnor2010}. We include in the reaction rates absorptions of $\nu_e$ and $\bar\nu_e$ by neutrons and protons, elastic scattering of all neutrinos on neutrons, protons, alpha particles, and heavy nuclei, and emission of $\nu_x$ due to $e^+e^-$ pair annihilation and nucleon-nucleon Bremsstrahlung. Inverse reactions are all calculated assuming Kirchoff's law. We currently ignore pair processes for electron type neutrinos, inelastic scattering, and all types of neutrino oscillations. Neutrino emissivities $\eta$, absorption opacities $\kappa_a$, and scattering opacities $\kappa_s$ are tabulated in 16 energy bins (logarithmically spaced up to 528\,{\rm MeV}), 82 density bins (logarithmically spaced between $10^{6}$ and $10^{15.5}$ g/cc), 65 temperature bins (logarithmically spaced between $0.05$ and $150\,{\rm MeV}$) and 51 values of the electron fraction $Y_e$ (linearly spaced between $0.01$ and $0.6$). 

To handle the densest/hottest regions of neutron stars, our Monte-Carlo algorithms makes two notable approximations. In regions with large scattering opacity ($\kappa_s \Delta t \geq 3$), we do not perform each scattering event individually, but instead directly move packets to a location drawn from the solution of a diffusion equation matching the outcome of individual scatterings in the limit of many neutrino-matter interactions~\cite{Foucart:2017mbt,Foucart:2018gis}. In regions with large absorption opacities ($\kappa_a \Delta t \geq 0.5$), we reduce the absorption opacity while keeping the total opacity $\kappa_a+\kappa_s$ and the equilibrium energy density $\eta/\kappa_a$ constant~\cite{Foucart:2021mcb}. This maintains the expected diffusion rate of neutrinos through the star and their equilibrium energy density, but effectively increase the equilibration timescale of neutrinos from $\kappa_a^{-1}$ to $(2\Delta t)$. This method, inspired by implicit Monte-Carlo algorithms~\cite{1971JCoPh...8..313F}, has the dual advantage of avoiding stiff coupling between the neutrinos and the fluid and of avoiding large numbers of emissions / absorptions during each time step. We have verified that for the short timescales considered here and the fluid configurations found in merger remnants, this approximation likely introduces errors that are significantly smaller than our current numerical errors~\cite{Foucart:2021mcb}. We also note that this approximation is not used much in the simulations presented in this manuscript, as the binaries evolved in this work rapidly form a black hole-disk system in which nearly all cells on our computational grid are optically thin to neutrinos (even if the disk in its entirety is not). This type or remnant is significantly easier to evolve using Monte-Carlo methods than the neutron star-disk remnants considered in our first Monte-Carlo simulation~\cite{Foucart:2020qjb}. A detailed discussion of our Monte-Carlo algorithm can be found in~\cite{Foucart:2021mcb}. 

In the simulations presented here, we use $10^6$ neutrino packets per species up to the collision of the neutron stars, and $4\times 10^7$ neutrino packets per species ($10^7$ at low resolution) thereafter. We note that this high number of packets is only really needed around the time of the collapse of the remnant to a black hole. We have also been able to evolve the post-merger remnant (after black hole formation) with as few as $10^6$ packets, due to the relatively low neutrino luminosity of the remnant and the fact that it is optically thin to neutrinos. However, after merger our numerical
grid is large enough that evolving more packets does not significant impact the cost of the simulations (we use $5\times 10^6$ grid cells in our post-merger evolution, and each fluid cell is significantly more expensive to evolve than a MC packet).

During the collapse to a black hole, on the other hand,
a high number of packets is actually crucial to the stability of the evolution, at least with the distribution of packets currently implemented in the SpEC code \footnote{It might be reasonable to simply ignore neutrino interactions in the very high density, hot matter formed during collapse to a black hole instead, but we have not so far attempted to do this}. With those methods, the simulations are unstable with $(10^6,4\times 10^6)$ packets. With $10^7$ packets, shot noise remains visible in the composition of the low-density regions, and the average $Y_e$ of the outflows increases noticeably (see results). With $4\times 10^7$ packets, shot noise is only observed close to the black hole, and the composition of the outflows largely matches expectations for a cold tidal tail unimpacted by neutrinos. We note however that the mass, temperature, and structure of the post-merger accretion disk and the other properties of the matter outflows are not notably impacted by the choice of the number of MC packets, as long as the simulations are stable (i.e. the simulations with $10^7$ and $4\times 10^7$ packets differ in those quantities by less than the difference between simulations with/without viscosity, or at different numerical resolution).

With $10^{6-7}$ packets per species across the entire computational grid, we do not have enough packets to measure the distribution function of neutrinos at a given time and position. However, this is not required for black hole-disk systems: neutrino-matter interactions act on relatively long timescales in these systems, and the Monte-Carlo code can thus rely on time averages of neutrino-matter interactions to capture the evolution of the fluid temperature and composition due to coupling of the fluid with neutrinos. For example, if we consider neutrino-matter interactions in the matter outflows, each grid cell in the polar region covers a fraction $\sim 10^{-4}$ of the angular spread of the outflows, as seen from the remnant. In these optically thin remnants, each neutrino packet also takes $\sim 0.5\,{\rm ms}$ to escape the grid, while the unbound matter itself is about $10$ times slower. Unbound matter escaping the grid will thus interact with $\sim 10^{3-4}$ neutrino packets as it moves from the remnant to the grid boundary. Given the low absorption opacities in the outflows, this likely correspond to only a small number of packets being actually absorbed -- but our code is designed to calculate energy, momentum, and lepton number exchanges between neutrinos and the fluid using the expectation value of neutrino-matter interactions for each packet (i.e. if a packet crosses an absorption optical depth $\tau=0.01$, it will deposit $1\%$ of its energy in the fluid, even if it only has a $1\%$ chance of being removed from the simulation). Accordingly, it makes use of all $\sim 10^{3-4}$ packets in the determination of the effects of neutrinos on the evolution of the fluid. Shot noise in neutrino-matter interactions will not be negligible, but it should be limited to a few percents relative errors for most of the matter leaving the grid as soon as a black hole is formed \footnote{In visualizations of $Y_e$ in Sec.~\ref{sec:results}, shot noise is only visible close to the black hole and in the atmosphere; this can be contrasted with the slightly noisier $Y_e$ observed in e.g. Fig.2 of~\cite{Foucart:2020qjb}, when most of the MC packets are used to resolve diffusion of neutrinos close to the neutron star surface. As mentioned above, however, these estimates are not valid during collapse, because a majority of MC packets are then used to capture the distribution of neutrinos in the hot, dense regions of the collapsing remnant.}

\subsection{Numerical grids}

The numerical grids used in our simulations before collapse of the remnant to a black hole are very similar to those chosen in our previous Monte-Carlo simulation~\cite{Foucart:2020qjb}. On the pseudospectral grid, the region around each neutron star is initially covered by a ball surrounded by 10 spherical shells, with the shells around the neutron star of mass $m_1$ having outer radius $r_1 = 0.7(m_2d)/(m_1+m_2)$, and the shells around the neutron star of mass $m_2$ having outer radius $r_2 = 0.7(m_1d)/(m_1+m_2)$. Here, $d$ is the separation between the two neutron stars, which remains constant in the coordinates of our numerical grid (as the grid rotates and shrink during the evolution of the binary to keep the neutron stars nearly stationary on the grid). 32 spherical shells centered on the center-of-mass of the binary also cover the region between $3d$ and $40d$. In between these shells, 14 filled cylinders cover the axis connecting the two neutron stars, and 12 distorted cylinders connect to the outer shells. Except for the use of a central ball covering the neutron star centers and the higher number of shells used around each neutron star, this domain decomposition is identical to what is used for binary black hole simulations in SpEC~\cite{Szilagyi:2009qz}. When the maximum baryon density on the grid rises more than $3\%$ above its initial value, we switch to a ``merger'' grid. A grid of $5^3$ cartesian blocks covers the forming remnant, building a cube with $\sim 40\,{\rm km}$ sides. 12 distorted cubes connect that central region to the outer shells covering the wave zones. After formation of an apparent horizon, our grid structure becomes much simpler, with 14 spherical shells covering the region between the horizon and the outer shells. The number of collocation points used within each cube/shell/ball/cylinder is chosen adaptively to match a target accuracy. Before switching to the merger domain, we require $10^{-4}$ relative errors in the spectral expansion of the metric in the wave zone, smoothly transitioning to $10^{-6}$ at the center of each neutron star. At later times, the target error is $10^{-4}$ everywhere. At high resolution, all target errors are mulitplied by $0.64$ (to match expected second order convergence on the finite volume grid). 

The finite volume grid is chosen so that there are 90 cells across the diameter of the star at the beginning of the simulation (112 cells at high resolution). As the grid shrinks during the inspiral of the binary, the effective resolution continuously increases during our simulation. Whenever the grid spacing in the lab frame decreases by $20\%$, we interpolate onto a new grid matching the initial grid spacing. Up to contact, all non-vacuum regions of our computational domain use this resolution, while vacuum regions are excluded from the finite volume grid (the code adaptively adds/removes small cartesian boxes to match the current location of the fluid). After formation of an event horizon, we use four nested cartesian grids centered on the black hole, each with $200^3$ live cells. The finest grid is a cube with side length of $40\,{\rm km}$ (in the arbitrary coordinate system that our simulation evolves into), with each coarser grid doubling the grid spacing. The outer boundary of the finite volume grid is thus $\sim 160\,{\rm km}$ from the center of the black hole, and the fastest outflows thus start leaving the computational domain $\sim 1.5\,{\rm ms}$ after collapse of the remnant to a black hole. In between contact and collapse, the finest grid is twice as long along the axis connecting the center of the merging neutron stars ($\sim 80\,{\rm km}$ long), and uses the same grid spacing as during the inspiral. For parallelization, each cube of $200^3$ cells is devided into $8^3$ subdomains ($160^3$ cells at low resolution). With $6$ ghost zones in each dimension (3 on each side), each subdomain thus has $31^3$ points. The $4^3$ central subdomains of the coarser grids are removed from the evolution domain, as they fully overlap a finer grid. Overall, this results in about $55$ millions total grid cells, $29$ millions total live cells (excluding ghost zones), and $1856$ subdomains (a larger number of sudomains would allow the code to scale to more cores, but also increase the ratio of ghost cells to live cells; the choices made here are a compromise between scalability and total cost of the simulations).

\subsection{Ejected matter}
\label{sec:ejecta}

Determining the eventual fate of the matter leaving the grid of a merger simulation is unfortunately not straightforward. Particles following geodesics of a time-independent metric conserve $u_t$, the time component of the velocity one-form, and are unbound if $u_t<-1$, with asymptotic Lorentz factor 
\beq
W_\infty=-u_t.
\eeq
For merger outflows, however, pressure gradients, out of equilibrium nuclear reactions (mainly r-process nucleosynthesis), and the associated neutrino emissions may all play an important role. An approximate method to account for pressure gradients and the release of energy by nuclear reactions is to assume conservation of $hu_t$ (Bernoulli), with $h=(u+P)/\rho_0$ the specific enthalpy. This is however strictly valid only along a streamline of a steady-state flow, and neglects neutrino cooling. The associated condition for matter to be unbound is $hu_t < -h_\infty$, with $h_\infty$ the expected asymptotic enthalpy (i.e. $h$ at low density, temperature, and for the composition of the fluid at the end of the r-process). We note that the value of $h_\infty$ is significantly impacted by the choice of reference baryon mass discussed in Sec.~\ref{sec:eos}, due to the division by $\rho_0$. Under these assumptions, we have
\beq
W_\infty = -hu_t/h_\infty.
\eeq
For the tables used in SpEC, $h_\infty\sim 1$ for the SFHo equation of state, and $h_\infty\sim 0.992$ for the LS220 equation of state (roughly the difference between taking as reference mass the mass per nucleon of a high-mass nucleus, or the mass of a neutron).

For equations of state that assume that the fluid is at its equilibrium composition, the first definition of $W_\infty$ typically underestimates the mass and velocity of the outflows (by neglecting their internal energy), while the second overestimates the mass of the outflows (by assuming that all of the internal energy is locally transformed into kinetic energy). For composition-dependent equations of state, the second estimate is even more of an overestimate because it effectively assumes that all of the energy released by nuclear reactions as the composition of the matter changes from neutron-rich to its composition at the end of the r-process ($Y_e\sim0.38$) is transformed into kinetic energy. In~\cite{Foucart:2021ikp}, we derived a third approximation to $W_\infty$ that accounts for the main deviation from this assumption: the loss of about half of the energy released by nuclear reactions through neutrino emission. Then
\beq
W_\infty \approx -\frac{hu_t}{h_\infty} \left(0.9968+0.0085 Y_e\right)
\eeq
which assumes that $45\%$ of the energy released by the r-process is lost to neutrinos. The numerical coefficients in this expression are taken from~\cite{Desai:2018rbc},  using estimates of the energy released by the r-process computed with the SkyNet code~\cite{Lippuner2015}. A more complex estimate of $W_\infty$ that accounts for the finite time required for the r-process to complete is also derived in~\cite{Foucart:2021ikp} (model `E' of that manuscript). In this manuscript, we report results using the `$u_t$' criteria, the `$hu_t$' criteria, and the model accounting for r-process heating, neutrino cooling, and the finite time needed for energy deposition. The latter model is used by default as our `best guess' model when we study the properties of the ejecta in more detail.

Simulations following the ejecta over weeks timescale have shown that, when these estimates disagree, none of them are particularly reliable when it comes to predicting the distribution of velocity or the geometry of the outflows, but the more advanced model does perform better as far as predictions for the total mass and total kinetic energy of the outflows are concerned~\cite{Darbha:2021rqj,Foucart:2021ikp}. We will see that our simulations show non-negligible differences between these various predictions. 

\subsection{Lagrangian tracers}
\label{sec:tracers}

In order to follow the evolution of individual fluid elements and provide a history of their evolution, as needed for example to feed our results into nuclear reaction network codes and to study out-of-equilibrium nuclear reactions in the matter ejected by the merger, we add Lagrangian tracers to the LS220-178-106 simulation (for the simulations using viscosity). We initialize tracers immediately before the collision of the two neutron stars, with each tracer representing $10^{-5}M_\odot$ of material. Each tracer keeps track of its grid position $x^i$, evolved at the end of each time step on the finite difference grid through the simple equation
\beq
\frac{\Delta x^i}{\Delta t}=v^i_T = \frac{u^i}{u^t}.
\eeq
The transport velocity $v^i_T$ is interpolated to the location of the tracer using 4$^{\rm th}$ polynomial interpolation. History files for each tracer include the density, temperature and composition of each tracer every $0.1\,{\rm ms}$, also interpolated to the location of the tracers using 4$^{\rm th}$ order methods. History for all tracer particles produced in these simulations is available on demand.

We note that due to different evolution methods, there is no guarantee that the tracers remain a reasonable sampling of the fluid over the entire duration of the simulation; this is only guaranteed to be true at the time at which the tracers are seeded. In order to assess deviations between the evolution of the tracers and the evolution of the fluid, we compare in this manuscript the properties of the tracers leaving the grid to the properties of the matter leaving the grid -- which should provide us with a reasonable estimate for how well the tracers sample the outflows.

\section{Results}
\label{sec:results}

\begin{figure*}
\includegraphics[width=0.24\linewidth]{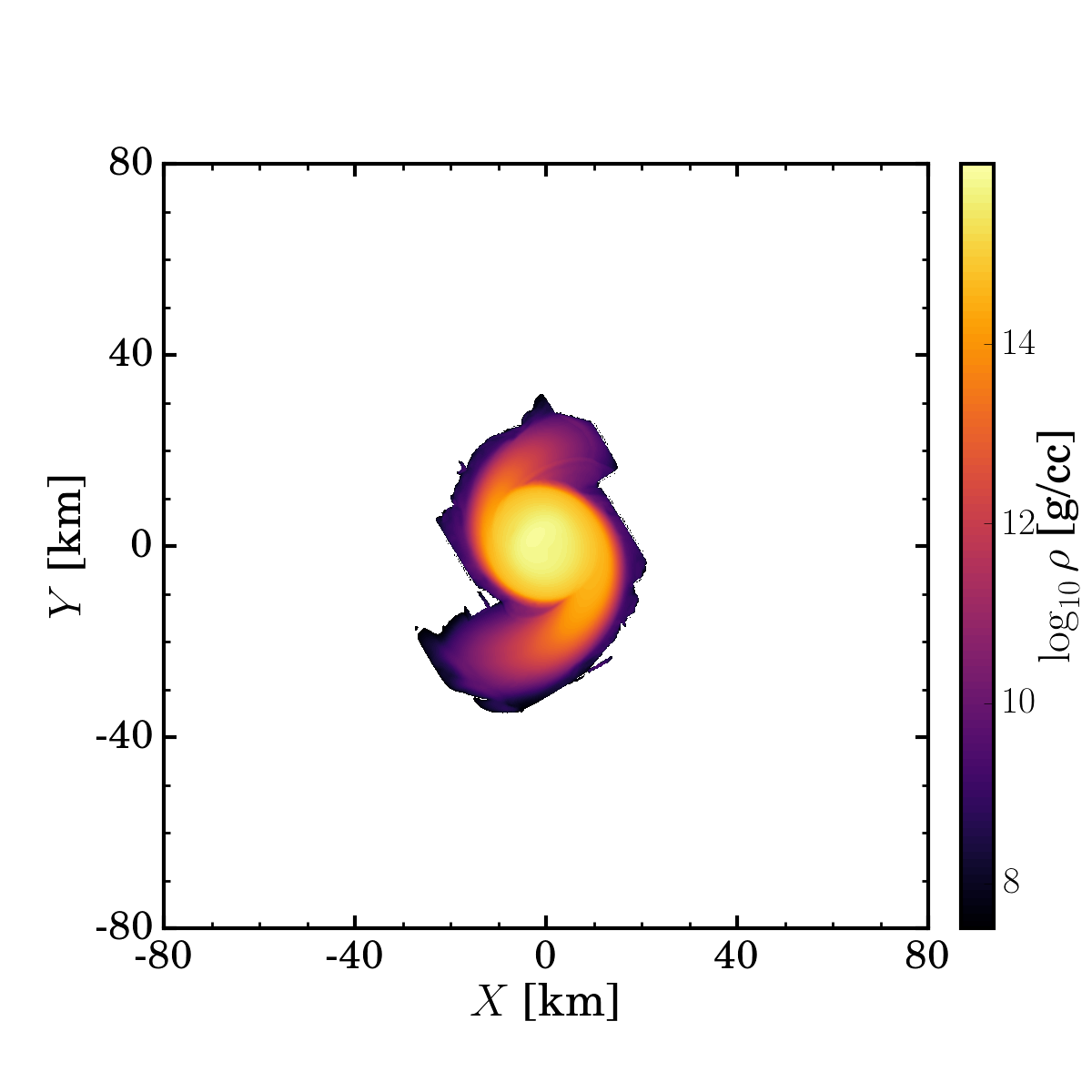}
\includegraphics[width=0.24\linewidth]{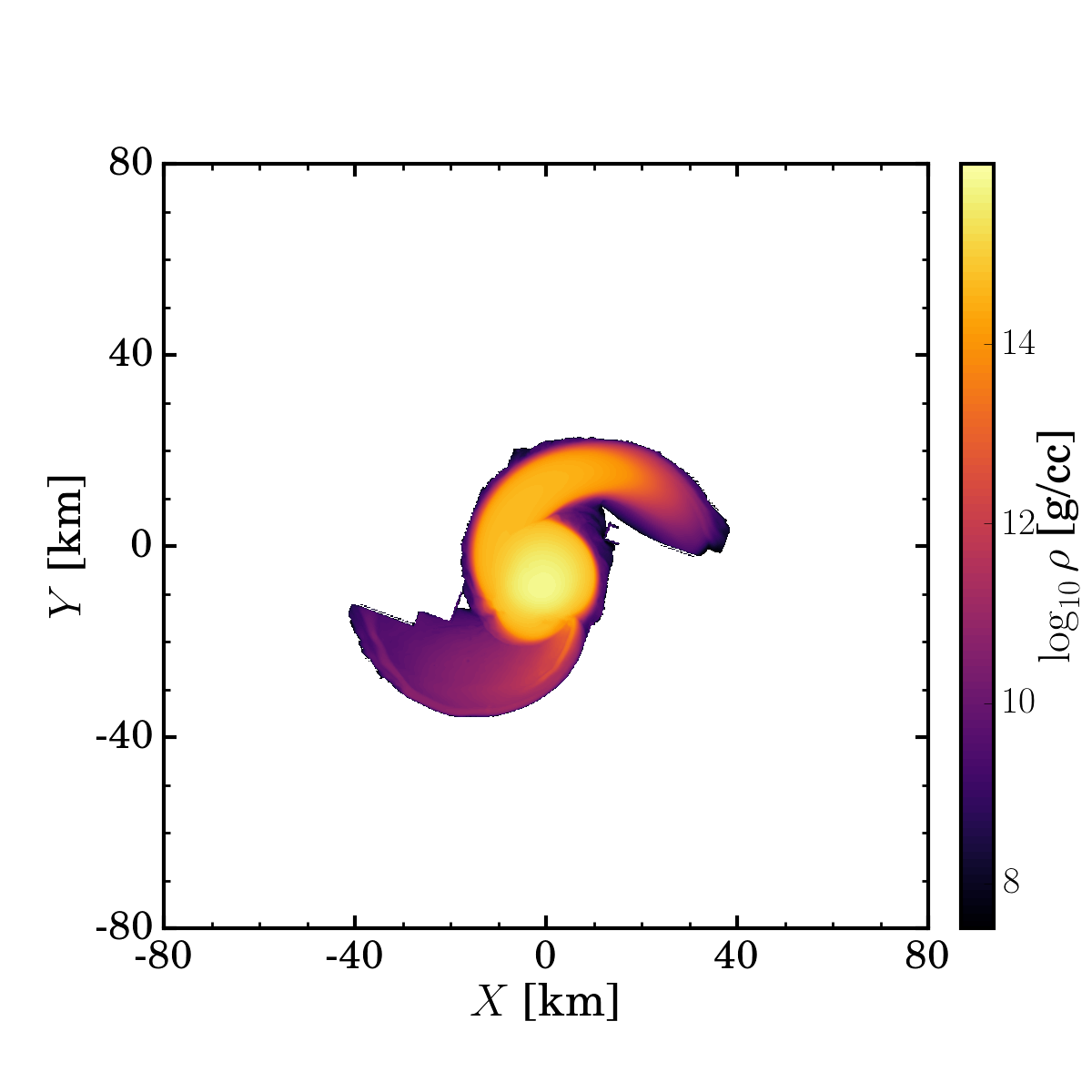}
\includegraphics[width=0.24\linewidth]{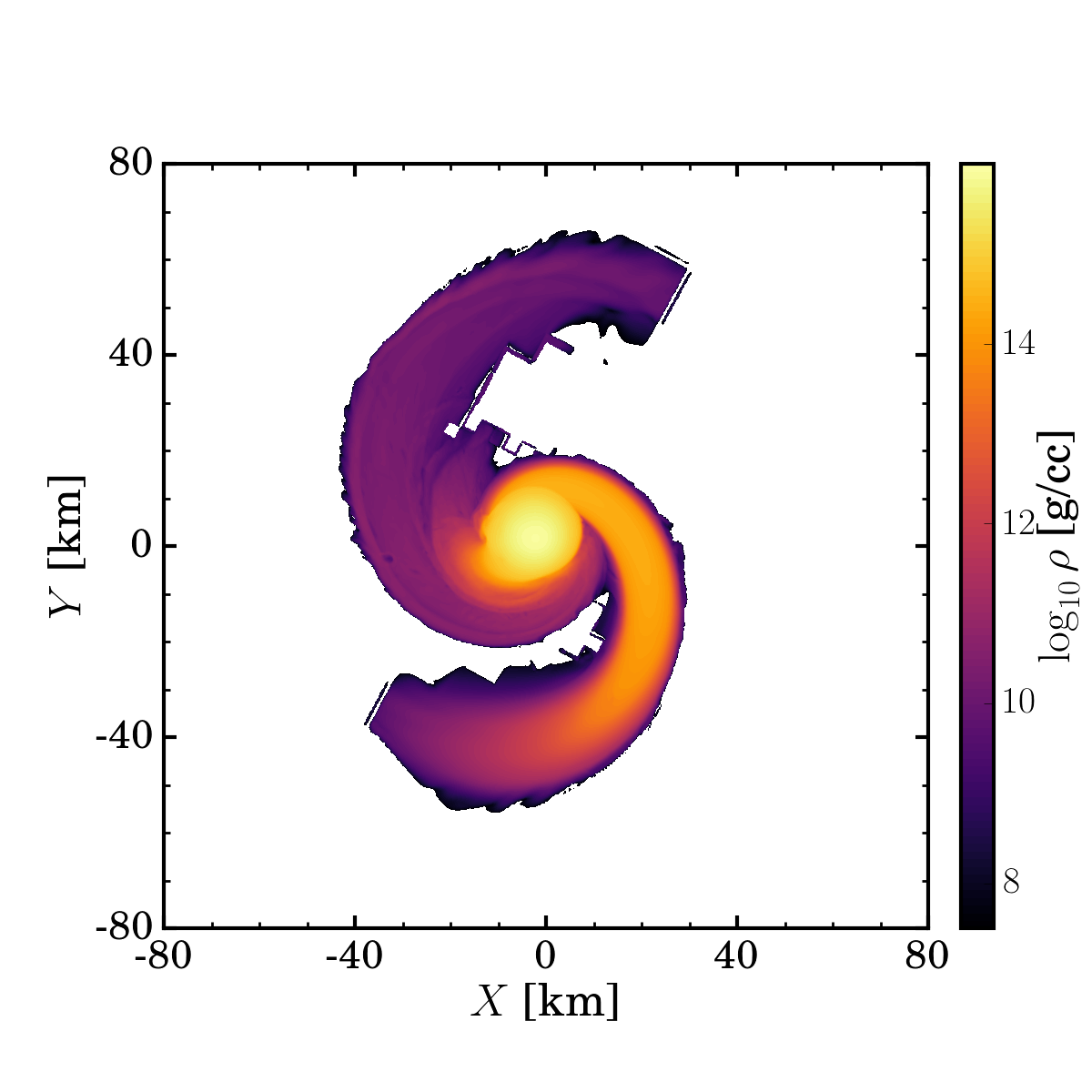}
\includegraphics[width=0.24\linewidth]{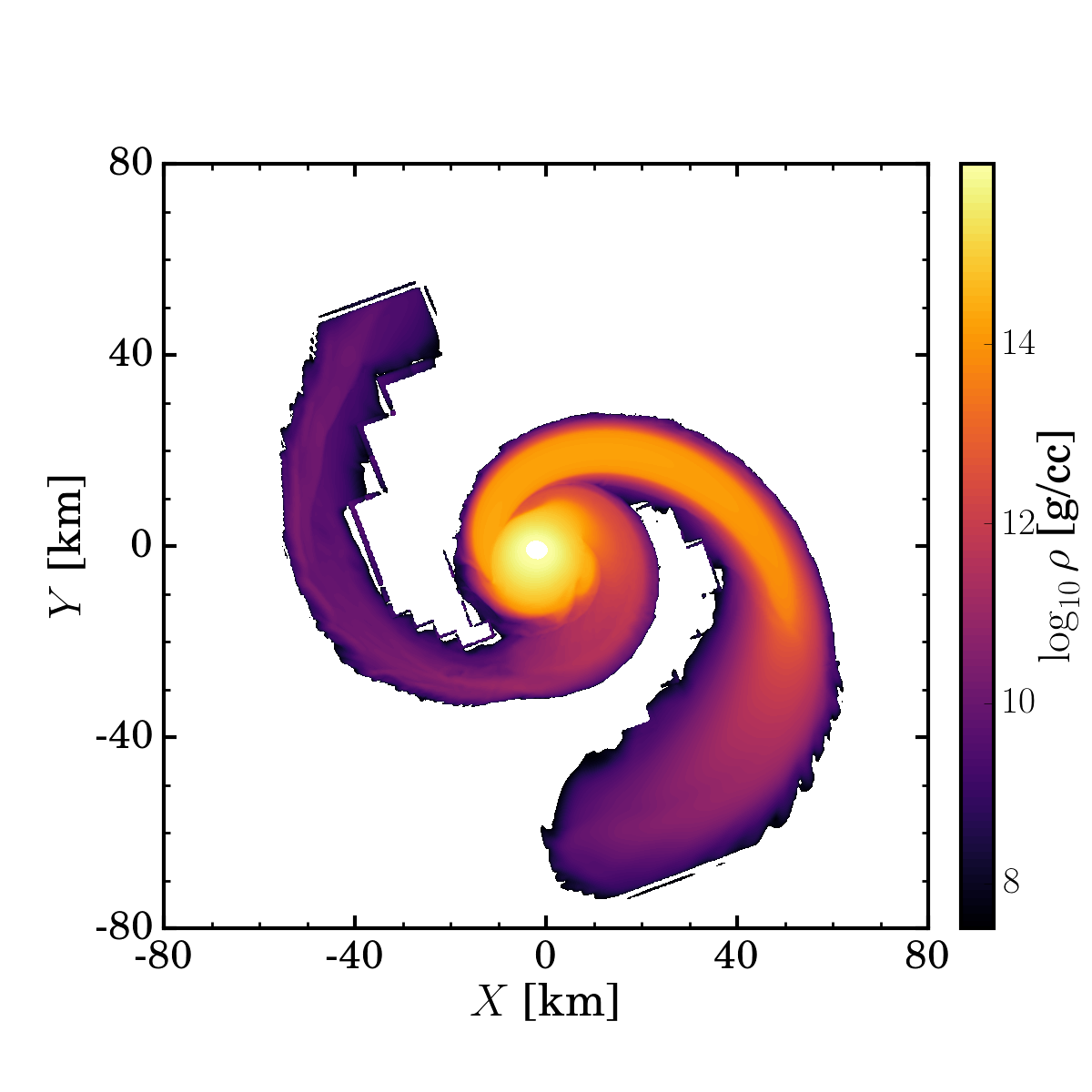}\\
\includegraphics[width=0.24\linewidth]{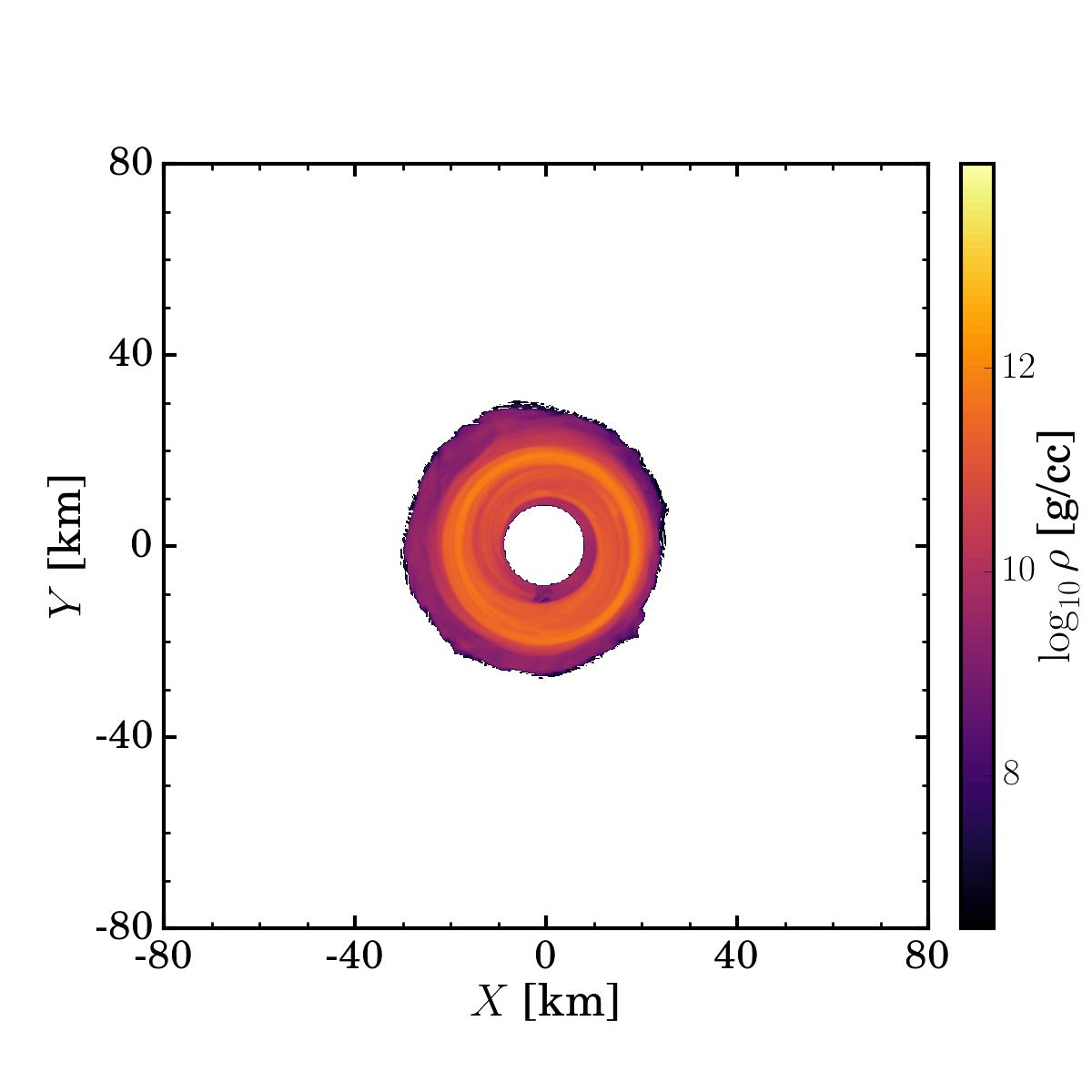}
\includegraphics[width=0.24\linewidth]{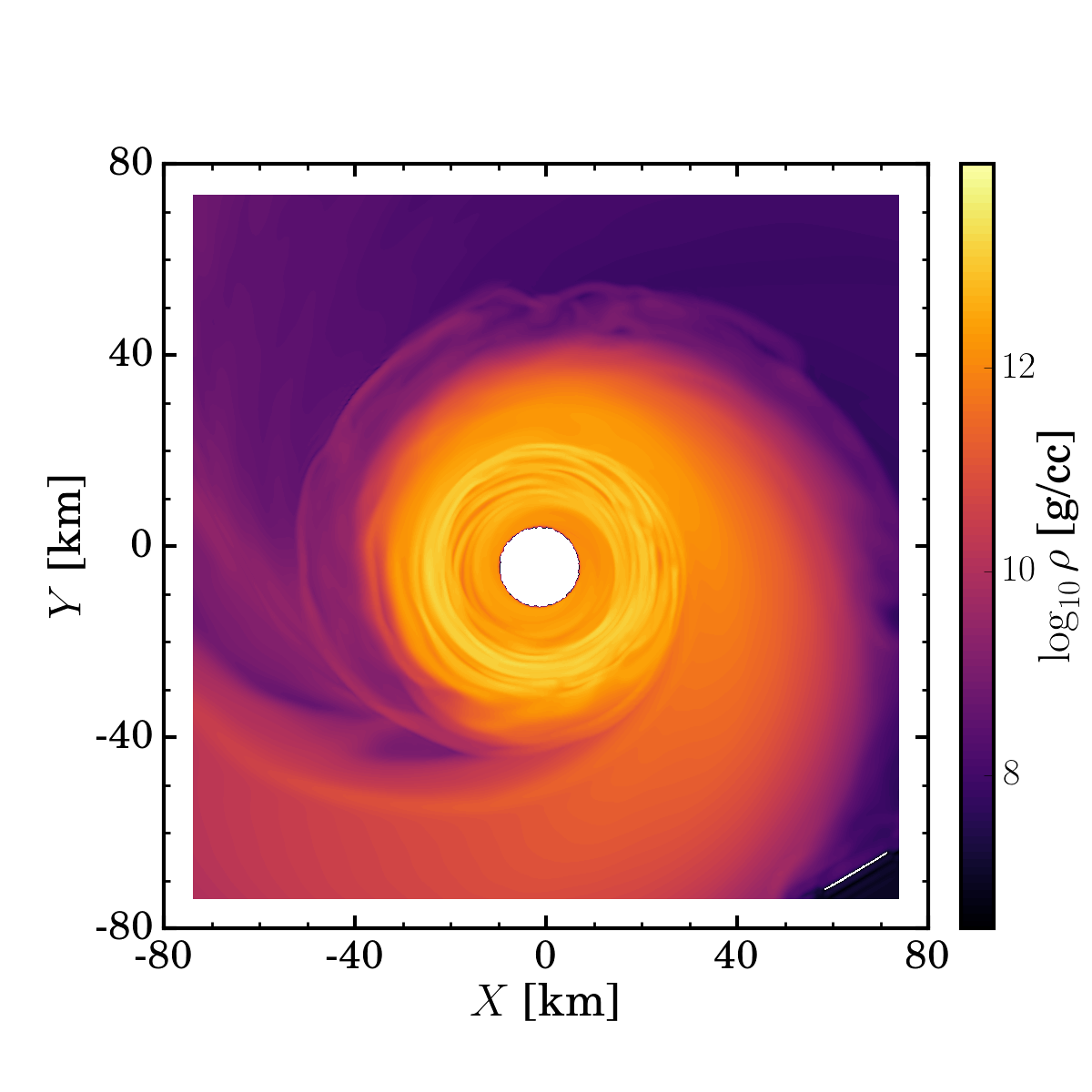}
\includegraphics[width=0.24\linewidth]{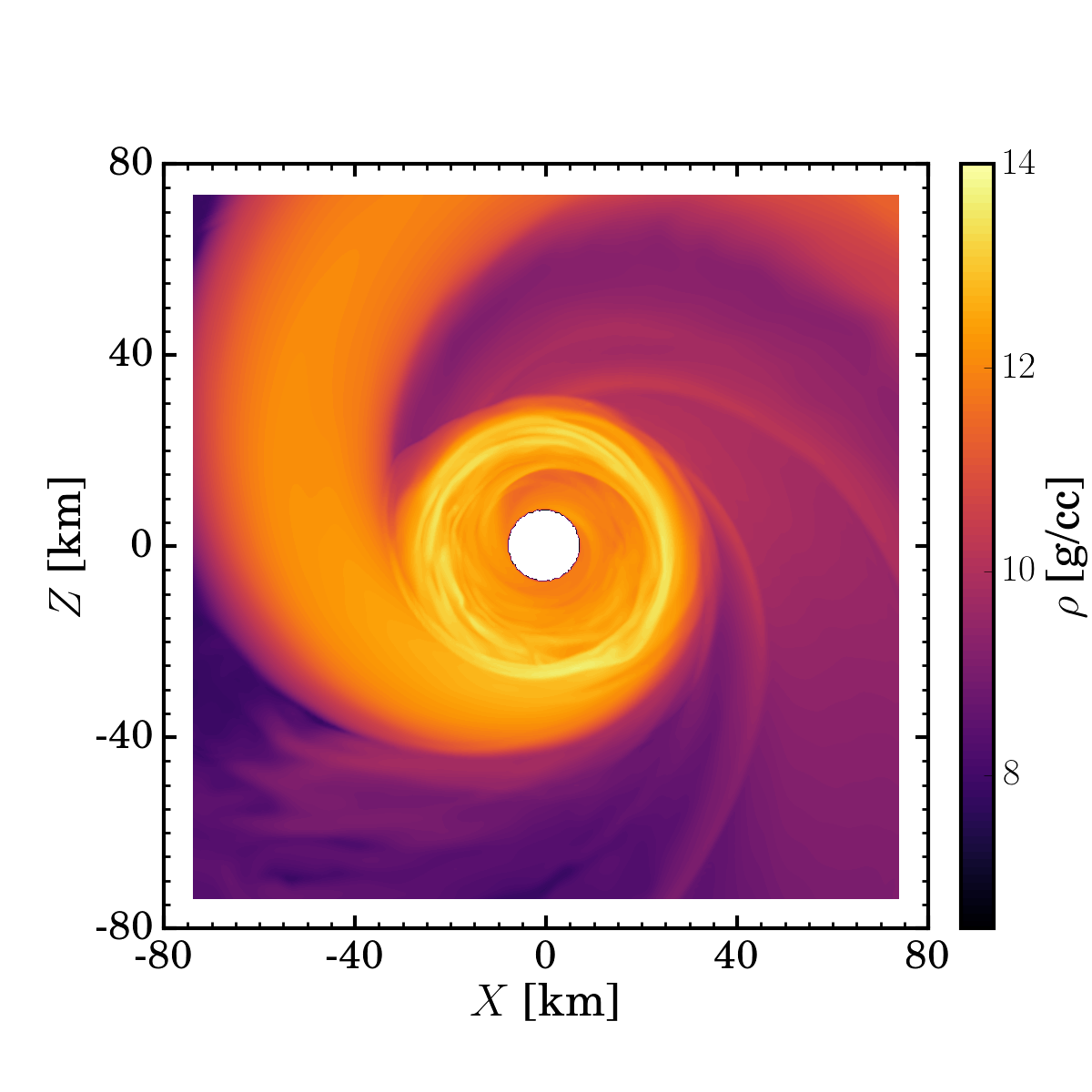}
\includegraphics[width=0.24\linewidth]{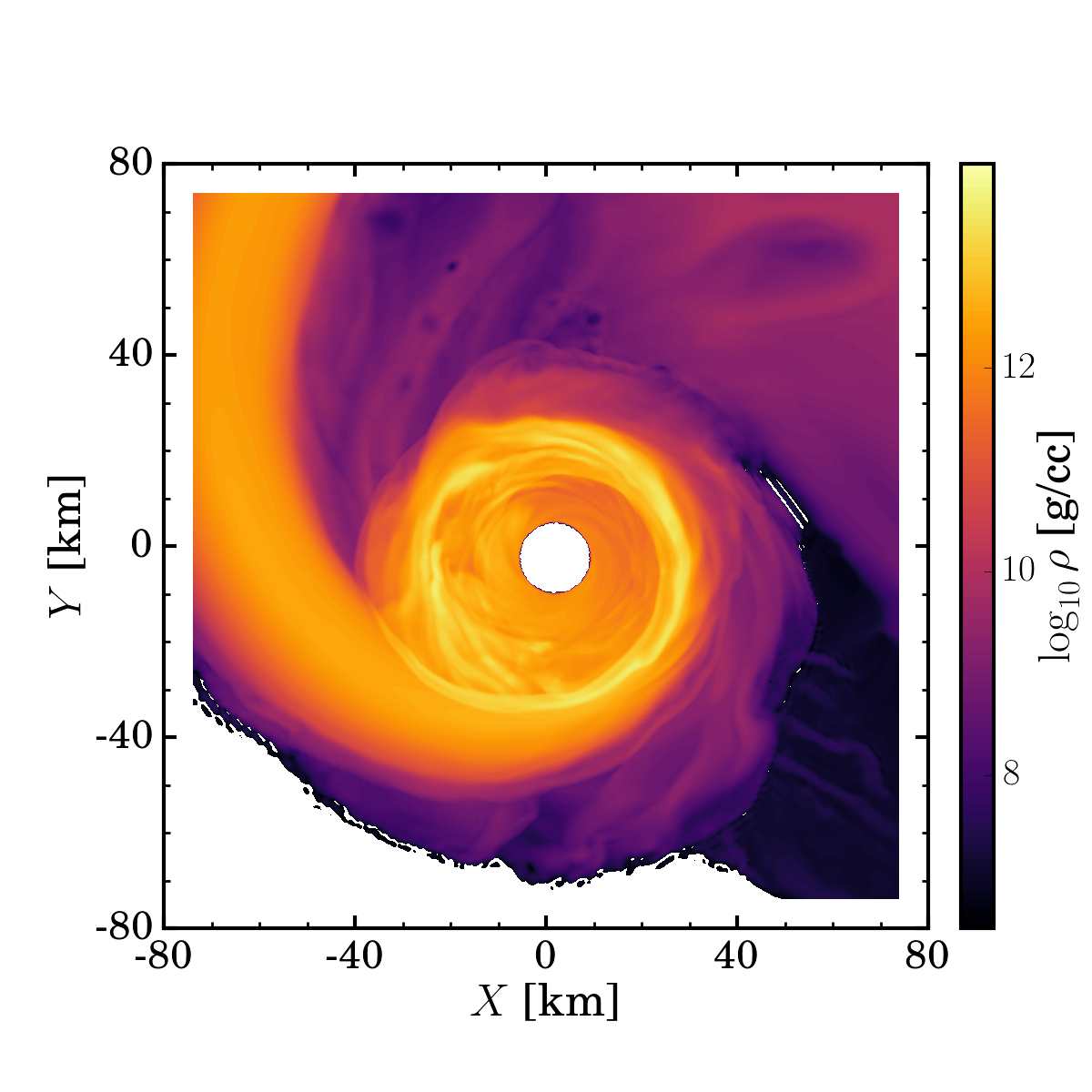}\\
 \caption{Baryon density in the equatorial plane of the binary just before an apparent horizon is detected (top), and $2\,{\rm ms}$ later (bottom). From left to right, we show simulations SFHo-161-151, SFHo-180-131, SFHo-178-106, and LS220-178-106. Note the different color scales on the top and bottom panels.}
\label{fig:Collapse}
\end{figure*}

\begin{figure*}
\includegraphics[width=0.8\linewidth]{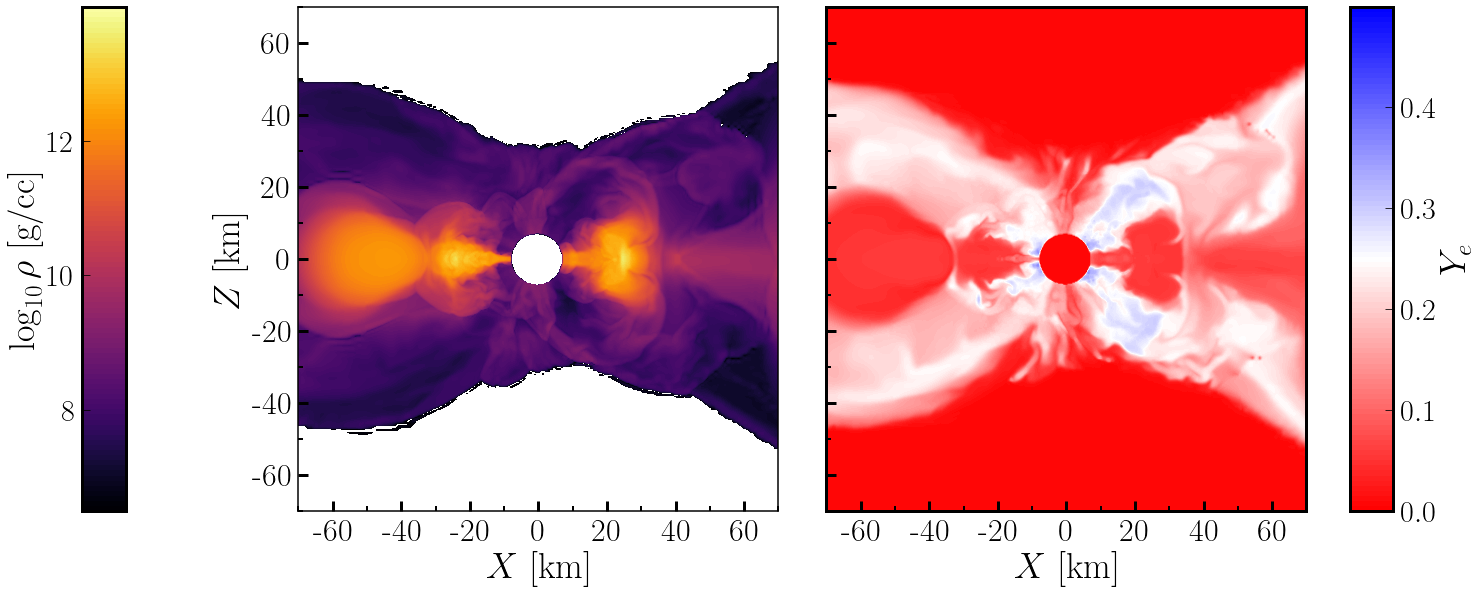}\\
\includegraphics[width=0.8\linewidth]{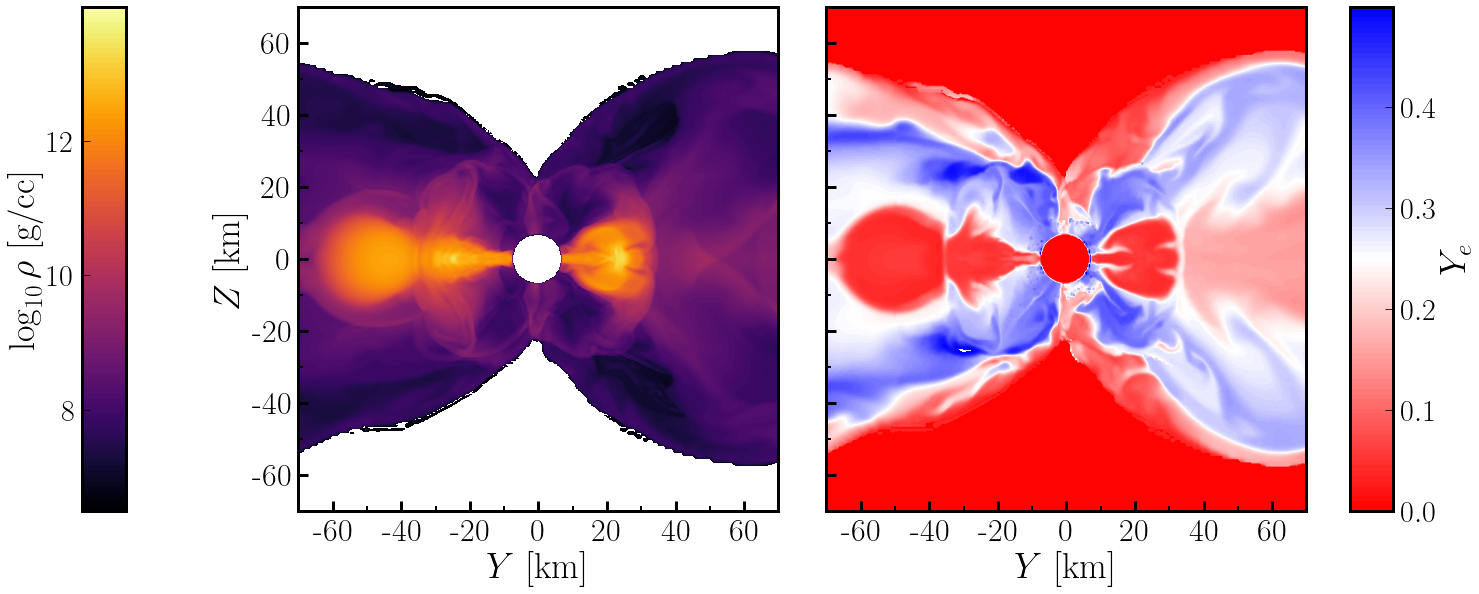}\\
 \caption{Baryon density and electron fraction of the fluid $2\,{\rm ms}$ after detection of an apparent horizon for simulations SFHo178-106 (top), and LS220-178-106 (bottom). We show the vertical slice $y=0$.}
\label{fig:CollapseVert}
\end{figure*}

\subsection{Disk and post-merger remnant}

The simulations performed in this work are all for high-mass systems. The lowest mass system has $M_{\rm tot}=2.84M_\odot$, well above the maximum mass of the LS220 and SFHo equations of state ($2.04M_\odot$ and $2.06M_\odot$, respectively). As a result, every merger results in the rapid formation of a black hole within a few milliseconds of the first contact between the neutron stars. We note that this is despite the fact that $2.84M_\odot$ is below the commonly used threshold masses for collapses~\cite{PhysRevLett.111.131101,PhysRevD.101.044006}, but in agreement with existing numerical results for such systems~\cite{Bernuzzi:2020txg}. This is most likely due to the fact that the threshold mass predictions are calibrated to more symmetric systems, as pointed out in~\cite{Bernuzzi:2020txg}. The density profile of the matter at the time at which an apparent horizon is first detected is shown on Fig.~\ref{fig:Collapse}. We can immediately see on this figure the major impact of the mass ratio on the structure of the neutron stars at merger. As expected, the near equal mass system shows only a very small amount of mass in the tidal tails that appeared on each side of the remnant. Higher mass ratio systems lead to the partial tidal disruption of the lower mass neutron star before collapse of the central object to a black hole. The equation of state also play an important role here, as less compact neutron stars are easier to disrupt. We see this for the $1.06M_\odot$ neutron star evolved with both the SFHo equation of state ($R_{\rm NS}=12\,{\rm km}$) and the LS220 equation of state ($R_{\rm NS}=12.8\,{\rm km}$): the latter is more strongly disrupted than the former. This is qualitatively visible in the larger extent of the tidal tail at merger (Fig.~\ref{fig:Collapse}, top panel), and quantitatively in the ejected mass discussed below.

\begin{table*}
\begin{tabular}{c|c|ccccc|cc}
Name & $M_{\rm bound}$ & $M_{\rm ej,u}$ & $M_{\rm ej,h}$ & $M_{\rm ej,r}$ & $\langle Y_{e,\rm ej} \rangle$ &  $\langle v_{e,\rm ej} \rangle$ & $M_{\rm BH}$ & $\chi_{\rm BH}$ \\
\hline
SFHo-161-151 & $5\, 10^{-4}M_\odot$ & $\lesssim 10^{-5} M_\odot$  & $\lesssim 10^{-5} M_\odot$ & $\lesssim 10^{-5} M_\odot$& NA & NA & $3.051M_\odot$ & $0.792$\\
SFHo-180-131 & $0.042M_\odot$ & $4.8\,10^{-5}M_\odot$ &$5.3\,10^{-5}M_\odot$  & $5.2\,10^{-5}M_\odot$ & NA &  NA & $3.014M_\odot$ & 0.774 \\
SFHo-178-106 & $0.137M_\odot$ & $0.0024M_\odot$ & $0.0052M_\odot$& $0.0034M_\odot$ & 0.045  & 0.133c &$2.665M_\odot$  & $0.696$\\
\hline
LS220-178-106-no-vis & $0.134M_\odot$ & $0.012M_\odot$ &$0.021M_\odot$  &$0.015M_\odot$ & $0.037$ & $0.144c$ & $2.650M_\odot$ & 0.683 \\
LS220-178-106-vis &  $0.126M_\odot$ & $0.011M_\odot$ & $0.020M_\odot$ & $0.014M_\odot$ & $0.038$ & $0.141c$ & $2.653M_\odot$ & 0.671\\
\end{tabular}
\caption{Remant properties measured $4\,{\rm ms}$ post-merger. We show the remaining bound mass $M_{\rm bound}$, and the ejected mass according to three different unbound criteria ($u_t<-1$ for $M_{\rm ej,u}$, $hu_t<h_{\rm min}$ for $M_{\rm ej,h}$, and the criteria from~\cite{Foucart:2021ikp} including r-process heating, neutrino cooling, and the impact of the finite time scale needed for heat deposition). For the latter criteria, we also show the average electron fraction and average velocity of the ejecta. Finally, we report the mass and dimensionless spin of the black hole. For the LS220-178-106 configuration, we show results with and without viscosity. The two mostly differ in the amount of mass in the remnant accretion disk,
and the spin of the remnant black hole.}
\label{tab:rem}
\end{table*}

In the near equal-mass case, most of the material outside of the initial apparent horizon does not have enough angular momentum to avoid falling into the newly formed black hole. Just $2\,{\rm ms}$ after black hole formation (see Fig.~\ref{fig:Collapse}, bottom panel), we are left with a very low mass accretion disk ($\lesssim 10^{-3}M_\odot$) around a rapidly spinning black hole ($\chi_{\rm BH}=0.792$). For this binary system, we do not measure any significant amount of ejecta in the simulation. This is the prototypical result for a high-mass, symmetric neutron star binary system. We will not discuss this system much more here, considering the lack of matter outside the black hole and thus the absence of interesting post-merger electromagnetic signals.

For the more asymmetric systems, there is enough matter with significant angular momentum pre-collapse to form a massive accretion disk and bound tidal tail around the newly formed black hole, and to unbind some material from the remnant (see Fig.~\ref{fig:Collapse}). The amount of bound material mostly varies with the mass of the merging neutron stars. It is nearly identical for SFHo-178-106 and LS220-178-106 ($\sim 0.15M_\odot$), but significantly lower for the more symmetric SFHo-180-131 ($\sim 0.04M_\odot$). As a result, the latter system ends up with a more rapidly spinning black hole ($\chi_{\rm BH}=0.77$) than the other two ($\chi_{\rm BH}\sim 0.7$). This is expected from the fact that the disk keeps a significant fraction of the angular momentum of the system in the SFHo-178-106 and LS220-178-106 simulations~\footnote{A rough Newtonian estimate $J_{\rm bound}/M_{BH}^2\sim (M_{\rm bound}/M_{\rm BH})(R_{\rm bound}c^2/(GM_{\rm BH}))^{1/2}$ shows that for $0.04M_\odot$ of material in a compact circularized disk around a $3M_\odot$ black hole, we expect $J_{\rm bound}/M_{BH}^2\sim 0.03$.}, but not in the SFHo-180-131 simulation.  The properties of the post-merger remnants are summarized in Table~\ref{tab:rem}. 
We can estimate the impact of viscosity and numerical resolution on these properties. We note $5\%-10\%$ differences in most quantities between viscous and non-viscous simulations. These differences in the microphysics used in the simulation are significantly larger than differences due to finite resolution. The disk mass, black hole mass and black hole spin only show variations of $(0.1-1)\%$ between the low and high resolution simulations. The only quantity in which we note a significant impact of our finite resolution is the average temperature of the fluid during and after collapse ($5\%-10\%$ differences, depending on the time of measurement).

\begin{figure}
\includegraphics[width=0.95\columnwidth]{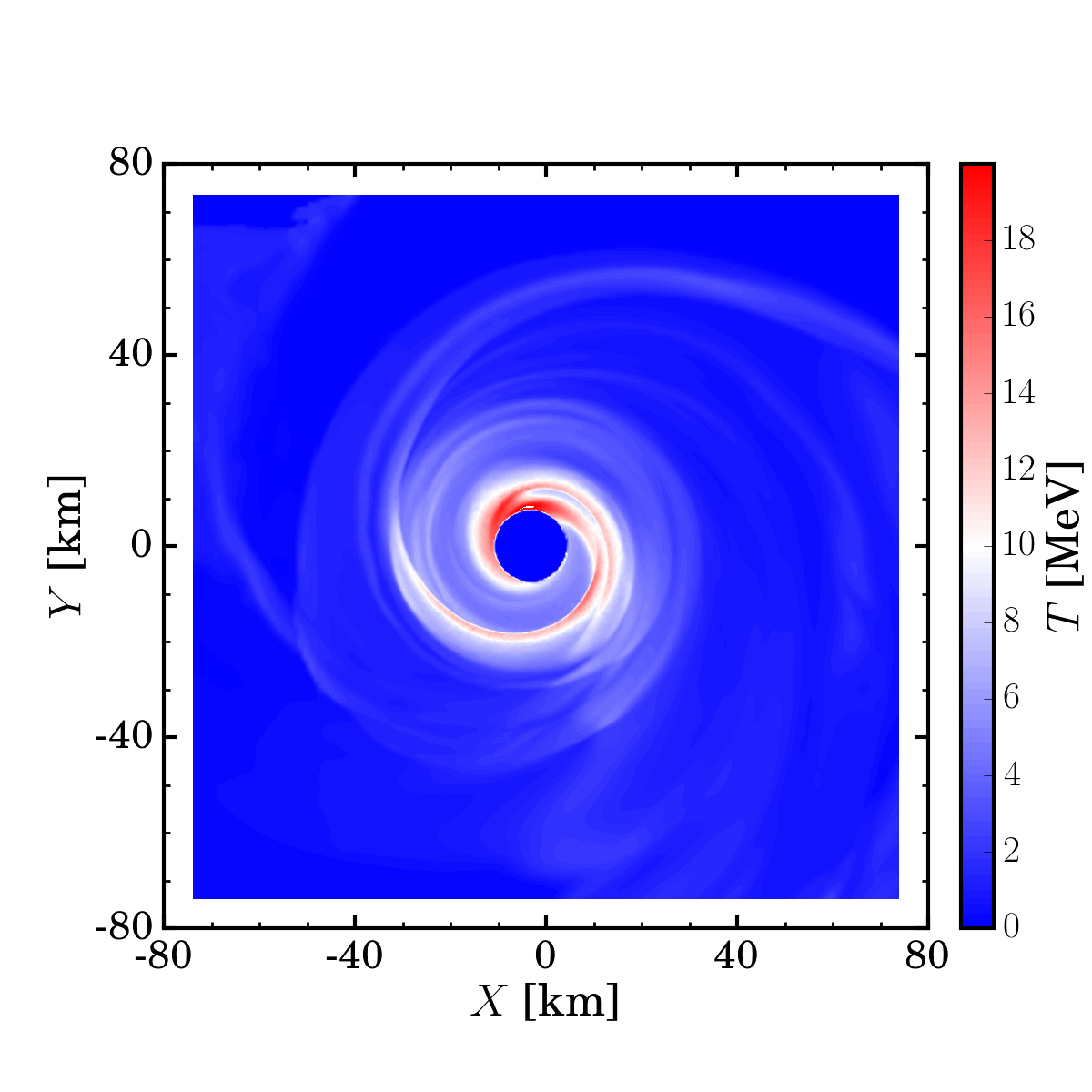}
 \caption{Temperature of the fluid in the equatorial plane $4\,{\rm ms}$ after black hole formation for simulation LS220-178-106. We clearly see the distinction between the hot disk $T\sim (5-20)\,{\rm MeV}$ and the cold extended tidal tail.}
\label{fig:Temp}
\end{figure}

\begin{figure*}
\includegraphics[width=\linewidth]{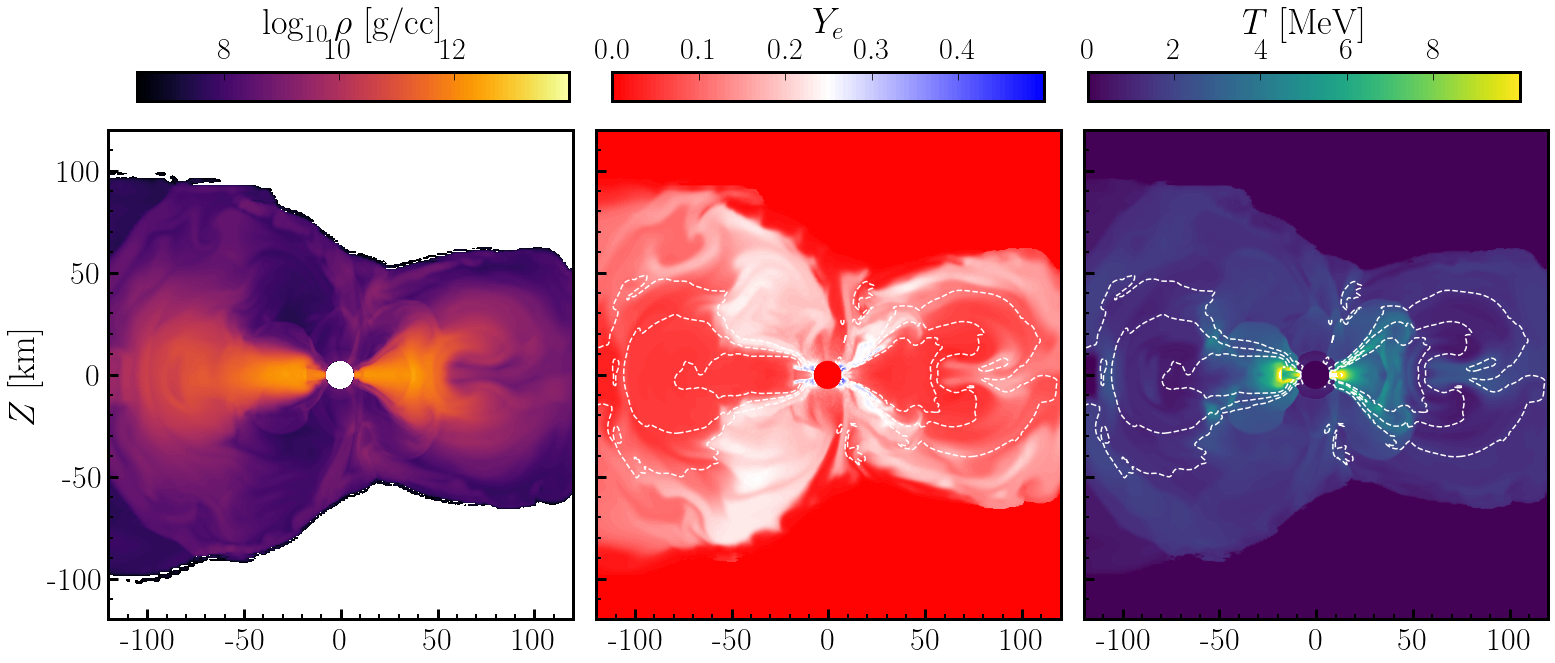}\\
\includegraphics[width=\linewidth]{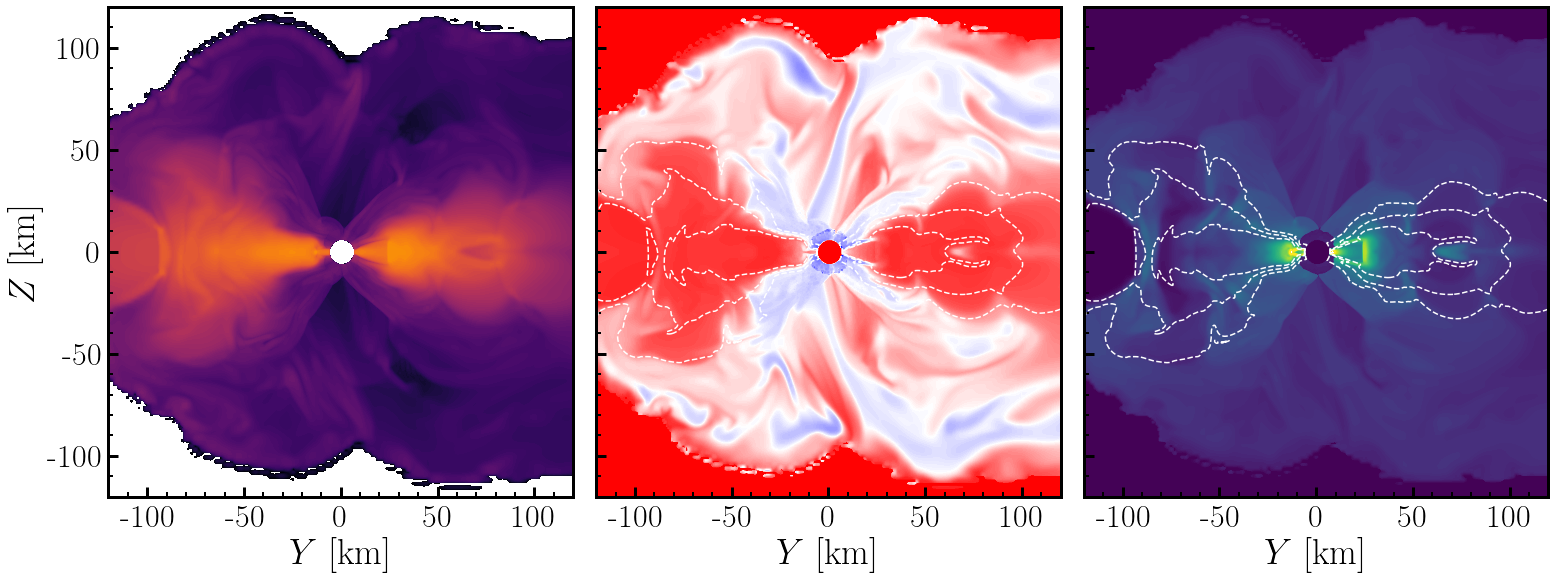}\\
 \caption{Vertical slice through the post-merger remnant $10\,{\rm ms}$ after collapse to a black hole. We show the SFHo-178-106 ({\it Top}) and  LS220-178-106 ({\it Bottom}) cases and the baryon density ({\it Left}), electron fraction({\it Center}), and temperature ({\it Right}). The structure of the
accretion disk is very similar in both simulations, but the LS220 simulation has higher electron fraction in the polar regions and a disk wind absent
at that time in the SFHo simulation.}
\label{fig:CollapseVert10ms}
\end{figure*}

For the more asymmetric systems a few milliseconds post-merger, most of the matter in $M_{\rm bound}$ is in the tidal tail, not the disk. For example, the LS220-178-106-vis system $4\,{\rm ms}$ post-merger only has $\sim 0.05M_\odot$ of material within $\sim 40\,{\rm km}$ of the black hole, where the matter appears to follow mostly circular orbits, and $\sim 0.09M_\odot$ of material in the bound tidal tail. This may be a particularly important distinction if we intend to predict late-time matter outflows, as the matter remaining outside of the black hole is a lot less bound than we might expect if we were to assume a compact accretion disk around the black hole remnant. The binding energy of the remnant disk has been shown to be an important predictor of the fraction of that matter that ends up ejected in post-merger outflows~\cite{Fernandez:2020oow}. As a result, SFHo-178-106 and SFHo-180-131 appear relatively similar $2\,{\rm ms}$ post-merger (Fig.~\ref{fig:Collapse}): their disks are not very different, but SFHo-180-131 has a much more massive tidal tail extending well out of the depicted region. We should expect SFHo-178-106 to produce more massive post-merger outflows both due to the higher amount of matter remaining outside of the black hole, and because of the more compact nature of the SFHo-180-131 remnant. The properties of the matter in the disk and tidal tail are also fairly distinct: while the disk is heated by shocks during its circularization, the tidal tail remains very cold (see Fig.~\ref{fig:Temp}). The same result holds for black hole-disk-tail systems in neutron star-black hole mergers; see for example Fig. 6 of~\cite{Brege:2018kii}, which is quite similar to our Fig.~\ref{fig:Temp} despite the different binary types, mass ratios, and equations of state used.

The vertical structure of the post-merger remnant $2\,{\rm ms}$ after formation of a black hole is shown on Fig.~\ref{fig:CollapseVert} for the most asymmetric binaries SFHo-178-106 and LS220-178-106. We see that most of the matter remains close to the equatorial plane, with no significant matter outflows in the polar regions. The system is beginning to form a moderately thick accretion disk, and remains extremely neutron rich in its densest region ($Y_e\lesssim 0.1$). The hotter disk corona has noticeably higher electron fraction ($Y_e\gtrsim 0.25$), especially for the LS220-178-106 configuration. Fig.~\ref{fig:CollapseVert10ms} shows a vertical slice through the same simulations $10\,{\rm ms}$ post-merger. By then, the disk has expanded to $\sim 50\,{\rm km}$ radius, with temperatures of a few MeV everywhere and continued accretion from the cold tidal tail (on the left of the figure). The innermost, hotter regions of the disk protonize to $Y_e\sim 0.2$ due to preferential emission of $\bar \nu_e$, while the rest of the disk remains more neutron rich. The polar regions are slightly more neutron rich than just after merger, presumably due to both mixing with neutron-rich material and absorption of $\bar\nu_e$, but they still have $Y_e\gtrsim 0.25$. The LS220-178-106 case shows a weak disk wind ($\sim 2\times 10^{-4}M_\odot$ of unbound material at that time, with $\langle Y_e\rangle =0.22$), while the SFHo-178-106 has no such features. While we do not continue the simulation beyond $10\,{\rm ms}$ with the numerical methods used in this manuscript, these late snapshots provide us with useful initial conditions for longer post-merger evolution of these systems. At the end of the evolution, both systems have $\sim 0.10M_\odot$ of bound material remaining.  Without viscosity, the remnant mass is $\sim 0.01M_\odot$ higher, but its average temperature is only $0.1\,{\rm MeV}$ higher ($2.6{\,\rm MeV}$ vs $2.5\,{\rm MeV}$) than when viscosity is included. The lack of temperature change is likely due to the self-regulating nature of the properties of the post-merger remnant: additional heating or cooling modifies the compactness of the disk, but not its temperature. A similar effect has been observed when including or ignoring neutrino cooling~\cite{Deaton2013}.

Systems SFHo-178-106 and LS220-178-106 were also evolved with a different numerical code and different microphysics (neutrino leakage, different prescriptions for the viscosity) in~\cite{Bernuzzi:2020txg}. Our resulting bound fluid masses agree within $\delta M_{\rm bound} = 0.02M_\odot$, the black hole masses within $\delta M_{\rm BH}=0.01M_\odot$ (correcting for the fact that~\cite{Bernuzzi:2020txg} reports the irreducible mass of the black hole, while we report its Christodoulou mass), and the dimensionless black hole spins agree to the two significant digits reported in~\cite{Bernuzzi:2020txg}. This is slightly larger than the variations observed between different resolutions of our own simulations, except for the black hole spin, but still extremely good agreement considering the significant differences between the two evolution algorithms, as well as differences in the exact definition of $M_{\rm bound}$. Overall, we thus see that the properties of the remnant black hole, post-merger accretion disk, and bound tidal tail are likely robustly predicted by both sets of simulations.

\begin{table*}
\begin{tabular}{c|ccccc|ccccc}
Name & $M_{\rm rem}$ & \cite{Kruger:2020gig} & \cite{Radice:2018pdn} &  \cite{Coughlin:2018fis} & \cite{Nedora:2020qtd} 
& $M_{\rm ej}$ &  \cite{Kruger:2020gig} &  \cite{Coughlin:2018fis} & \cite{Nedora:2020qtd} & \cite{Dietrich:2016fpt} \\
\hline
SFHo-161-151 & $5\, 10^{-4}M_\odot$ & $1.3 \, 10^{-6} M_\odot$ & $0.001 M_\odot$ & $0.001 M_\odot$ & $0.06 M_\odot$
 & $\lesssim 10^{-5} M_\odot$ & $0.0018 M_\odot$ & $0.0060 M_\odot$ & $0.0053 M_\odot$ & $0.0048 M_\odot$\\
SFHo-180-131 & $ 0.042M_\odot$ & $0.040 M_\odot$ & $0.025 M_\odot$ & $0.001 M_\odot$ & $0.22 M_\odot$
 & $5\,10^{-5} M_\odot$ & $0.010 M_\odot$ & $0.009 M_\odot$ & $0.0035 M_\odot$ & $0.014 M_\odot$\\
SFH0-178-106 &  $ 0.14M_\odot$ & $0.21M_\odot$ & $0.21 M_\odot$ & $0.001 M_\odot$ & $0.068 M_\odot$
 & $0.0034 M_\odot$ & $0.029 M_\odot$ & $0.039 M_\odot$ & $0.21M_\odot$ & $0.032 M_\odot$\\
LS220-178-106 &  $ 0.15M_\odot$ & $0.28M_\odot$ & $0.21 M_\odot$ & $0.023 M_\odot$ & $0.50 M_\odot$
 & $0.015 M_\odot$ & $0.030 M_\odot$ & $0.033 M_\odot$ & $88M_\odot$ & $0.0067 M_\odot$\\
\end{tabular}
\caption{Comparison between our simulation results and the prediction of existing analytical modes fitted to numerical simulations for the mass remaining outside of the black hole after merger ($M_{\rm rem}$) and the mass of the dynamical ejecta ($M_{\rm ej}$). We note that the analytical models are not fitted to simulations in this regime. Their poor performance is thus expected; we mainly provide this comparison as a caution against using existing models in the high total mass, asymmetric binary regime.}
\label{tab:fits}
\end{table*}

Finally, we can compare our results to the predictions of analytical fits to preexisting numerical simulations (Fig.~\ref{tab:fits}). We do not expect particularly good agreement there, as these fits are known to be inaccurate for high-mass systems~\cite{Camilletti:2022jms}, which lie outside of their range of applicability. We mostly provide those results as an additional warning against using such fits in this regime. For the mass remaining outside of the black hole after merger, $M_{\rm rem}=M_{\rm bound}+M_{\rm ej}$, we compare our results to~\cite{Kruger:2020gig,Radice:2018pdn,Coughlin:2018fis,Nedora:2020qtd}. The model of~\cite{Kruger:2020gig} is meant to capture high-compactness results and the qualitative overall behavior of binary mergers; the models of~\cite{Radice:2018pdn} attempts to use the reduced tidal deformability as its main parameter, to facilitate comparisons with GW observations. In~\cite{Coughlin:2018fis}, the authors focus on capturing the transition between collapsing and non-collapsing systems; and in~\cite{Nedora:2020qtd} the focus
was on robust fitting of the region covered by simulations with physics-agnostic polynomial functions. We see that two of the models with the most physics input (\cite{Kruger:2020gig,Radice:2018pdn}) capture the value of $M_{\rm rem}$ within a factor of two or better. The model from~\cite{Coughlin:2018fis} was developed before simulations at high total mass and asymmetric mass ratio showed that massive disks could be formed in such systems, and thus naturally underestimates the mass of the remnant. Finally, the polynomial fits of~\cite{Nedora:2020qtd} have larger extrapolation errors than the other models, and are unreliable in this regime (as pointed out in the original manuscript, and unsurprisingly for a physics-agnostic polynomial fit).

\subsection{Matter outflows}

We now turn our attention to the matter flagged as ejected during the merger. As we evolve the simulations until the unbound parts of the tidal tail have left the grid, its characteristics are determined from our sampling of the matter leaving the computational grid. Simulation SFHo-161-151 does not produce any significant amount of unbound material before the remnant collapses to a black hole, and simulation SFHo-180-131 produces such a small amount of unbound material ($5\times 10^{-5}M_\odot$) that its measured properties are likely significantly impacted by numerical errors. Accordingly, we focus our attention once more on the most asymmetric systems, SFHo-178-106 and LS220-178-106.

In Table~\ref{tab:rem}, we report the unbound mass measured according to the `geodesic' criteria $u_t<-1$, the `Bernoulli' criteria $hu_t<h_{\rm min}$, and the more complex criteria accounting for r-process heating and neutrino cooling described in~\cite{Foucart:2021ikp}. We see that the predicted unbound mass is significantly impacted by the choice of unbound criteria. The Bernoulli criteria nearly certainly overestimates the amount of kinetic energy given to the ejecta; we mostly report it here because it has been widely used in the literature (though mainly for hot dynamical ejecta). The more advanced criteria does however account for a real physical effect: the deposition into the outflows of a fraction of the energy released as r-process nucleosynthesis proceeds, and the matter becomes less neutron rich and forms more strongly bound nuclei (instead of mostly neutrons). This effect cannot be captured by numerical simulations that do not allow for out-of-equilibrium nuclear reactions in the outflows, even if they were to be evolved for multiple seconds (as they mostly evolve at constant $Y_e$ once weak interactions become negligible). Our most advanced estimate is expected to do a little better, though it still only qualitatively captures the impact of nuclear reactions. A full hydrodynamical evolution of the outflows with realistic nuclear heating would be required to really determine the properties of the outflows at the end of the r-process~\cite{Klion:2021jzr,Darbha:2021rqj}. In~\cite{Foucart:2021ikp}, we found that the more advanced model did reasonably well when attempting to predict the mass and total kinetic energy of the outflows, but not when attempting to predict the velocity distribution or overall geometry of the ejecta. As a result, while we here take the more advanced model including heating and cooling as our 'default' prediction for the properties of the ejecta, it is fair to say that differences between that model and the simple geodesic model may still be our safest bet when estimating uncertainties in our predictions. In this case, that uncertainty turns out to be about $10\%-50\%$ (and $\sim 25\%$ for the configuration with the largest amount of ejected mass). We note that this is a larger source of error than the numerical resolution or viscosity model used in the simulation. As for the post-merger remnant, viscosity impacts the outflow mass at the $5\%-10\%$ level, while numerical resolution impact the outflow mass at the $\sim 1\%$ level. We note that this is a surprisingly low difference between low and high resolution simulations, possibly due to the fact that nearly all of the ejecta is in a cold tidal tail, and easier to resolve than the hot, shocked ejecta observed in many other systems.

The equation of state used in our simulations plays a more important role in setting the ejected mass than it did in setting the properties of the disk, bound tail, and remnant black hole. The SFHo-178-106 simulation ejects $\sim 0.003M_\odot$ of material only, while the LS220-178-106 simulation ejects $\sim 0.014M_\odot$ (respectively $0.002M_\odot$ and $0.011M_\odot$ using the geodesic criteria). We can compare those results to the simulations of Bernuzzi {\it et al}~\cite{Bernuzzi:2020txg}. In that manuscript, results were reported for the geodesic criteria. The identical LS220 binary without viscosity evolved at different resolutions ejected $(0.0086-0.0105)M_\odot$ without viscosity, and the same system with viscosity ejected $(0.0084-0.0138)M_\odot$. The SFHo binary with viscosity ejected $(0.0014-0.0015)M_\odot$. We thus see fairly consistent results for the LS220 system (within errors due to finite resolution and differing microphysics). Differences between codes and between numerical resolutions are smaller than the uncertainty due to the choice of unbound criteria for LS220. For SFHo, there is nearly a factor of two difference between the two codes, but the total amount of mass ejected is much smaller, and the differences are thus likely explained by finite resolution errors in the simulations. 

As for the disk mass, we can compare our results to the prediction of existing analytical formulae fitted on numerical simulations. Here, we consider the predictions of~\cite{Kruger:2020gig,Coughlin:2018fis,Nedora:2020qtd,Dietrich:2016fpt}. We see that the results are significantly worse than for the total mass left outside the black hole. The models of~\cite{Kruger:2020gig,Coughlin:2018fis,Dietrich:2016fpt} very roughly captures the dependence of the results in the neutron star masses at the qualitatively level, but are quantitatively off by orders of magnitude and do not capture the equation of state dependence. The model from~\cite{Nedora:2020qtd}, in this regime, extrapolates a prediction for $\log_{10}(M_{\rm dyn})$, and provides unphysical predictions for the most asymmetric systems. None of these models come close to the accuracy that would be required to provide predictions for kilonovae signals.

\begin{figure}
\includegraphics[width=0.99\columnwidth]{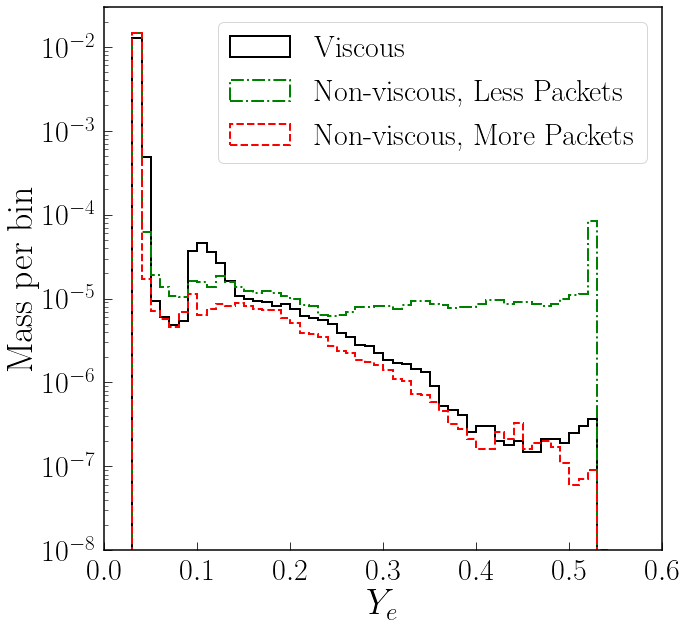}
 \caption{Distribution of $Y_e$ for the unbound matter of simulations LS220-178-106-vis (solid black) and LS220-178-106-novis (dashed red). In both cases, very neutron rich material with $Y_e\lesssim 0.05$ heavily dominates. We also show the result of a non-viscous simulation using a lower number of MC packets ($10^7$, dot-dashed green). Insufficient MC resolution during collapse leads to the ejection of a small amount of high-$Y_e$ matter.}
\label{fig:Ye}
\end{figure}

In all simulations that produce significant amounts of unbound material, that ejecta is in the form of a cold, neutron-rich, equatorial tidal tail. This is as expected for highly asymmetric mergers. We measure average electron fractions $Y_e\sim 0.04$, independent of the binary system or viscosity model. We note that this is one of the few differences between our results and those of Bernuzzi {\it et al}~\cite{Bernuzzi:2020txg}. In~\cite{Bernuzzi:2020txg}, the addition of a subgrid viscosity model led to a measurable increase in the average electron fraction of the outflows (to $Y_e\sim (0.06-0.08)$ for LS220-178-106). On Fig.~\ref{fig:Ye}, we show the electron fraction of our outflows for three different versions of the LS220-178-106 case. While there is certainly a slight protonization of the outflows in the viscous simulation (bump at $Y_e\sim 0.1$), that protonization is significantly smaller than in~\cite{Bernuzzi:2020txg}. Our simple viscosity model is expected to produce more heating than the model of~\cite{Bernuzzi:2020txg}, unless the corrections to the viscous stresses that we implemented in~\cite{Duez:2020lgq} to recover conservation of energy in the Newtonian limit end up playing a significant role in this process. Otherwise, the most likely source for this discrepancy would be neutrino physics and the neutrino transport algorithm. We discuss the evolution of the neutrinos in our simulation in more details in Sec.~\ref{sec:nu}, but note that undersampling of the MC packets during
merger can impact the protonization of the outflows. Comparing the simulations with $10^7$ packets and $4\times 10^7$ packets on Fig.~\ref{fig:Ye}, we see that the tail of the $Y_e$ distribution is underresolved in this simulation. We note that this is not overly surprising, as decreasing the number of packets during merger to even $4\times 10^6$ leads to unstable evolution of the collapsing remnant, in the form of a rapid rise of electron fraction and neutrino energy density before formation of a black hole. This part of the evolution is difficult for the MC code to resolve, with densities and mass-weighted average temperature rising to $10^{16}\,{\rm g/cc}$ and $200\,{\rm MeV}$ just before formation of an apparent horizon, and rapid changes in the density and temperature of the collapsing matter. Because these regions are causally disconnected from the matter remaining outside the black hole, their evolution does not impact the properties of the remnant and outflows -- but only as long as the large amount of MC packets needed to capture the evolution of the neutrinos in these regions during the last few microseconds of the collapse does not lead to increased sampling noise in the rest of the remnant, and the collapsing remnant itself can be stably evolve up to the time at which we find an apparent horizon. Undersampling of the MC packets is however unlikely to explain the differences between our results and~\cite{Bernuzzi:2020txg}, as it leads to an overestimate of $Y_e$ in our lower resolution simulations.

\begin{figure*}
\includegraphics[width=0.8\linewidth]{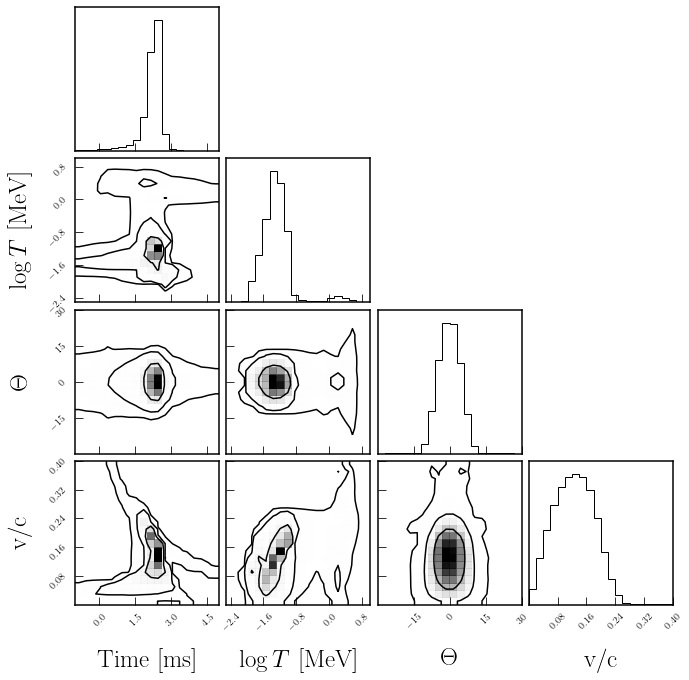}
\label{fig:cornervis}
\caption{Corner plot of the distribution of unbound matter leaving the grid as a function of time after the formation of an apparent horizon (estimate of the time at which the matter crosses a sphere $\sim 180\,{\rm km}$ from the black hole), temperature, polar angle, and velocity. We also show $(1,2,3)-\sigma$ contours of the distribution function in each 2D plot. The data in this plot is from simulation LS220-178-106. Plot generated using the corner python library from~\cite{corner}.}
\end{figure*}

Fig.~\ref{fig:cornervis} provide a more detailed view of the properties of the outflows for the LS220-178-106 binary. We see that most of the matter leaves the grid within $\sim 3\,{\rm ms}$ after black hole formation, with a broad distribution of velocity centered at $v\sim (0.10-0.15)c$. All of the matter is concentrated within $\sim 15^\circ$ of the equatorial plane, and very cold.

\subsection{Neutrinos}
\label{sec:nu}

In the rapidly collapsing post-merger remnants simulated in this manuscript, neutrinos do not play as much of a role as in systems creating long-lived neutron star remnants. During inspiral, neutrinos have a negligible impact, as usual in neutron star binaries. Heating of the outer layers of each neutron star due to tides, numerical viscosity, and subgrid viscosity does lead to the production of neutrinos of total energy $(10^{-4}-10^{-6})M_\odot c^2$, but these neutrinos are not dynamically important. During merger, the total energy of the neutrinos as measured by stationary observers (which may become unphysical close to the high density center of the collapsing neutron star core) quickly rises to $(0.005-0.04)M_\odot c^2$, only to immediately fall back to $(10^{-6}-10^{-5})M_\odot c^2$ within a few tens of milliseconds of the formation of an apparent horizon. The neutrino luminosity similarly shows a brief peak before the formation of an apparent horizon, a rise as a hot accretion disk starts forming around the remnant black hole, and finally steady emission from the remnant disk that should evolve over much longer timescales ($\sim 100\,{\rm ms}$). The total luminosity remains however much lower than for non-collapsing binaries. 

\begin{figure*}
\includegraphics[width=0.4\textwidth]{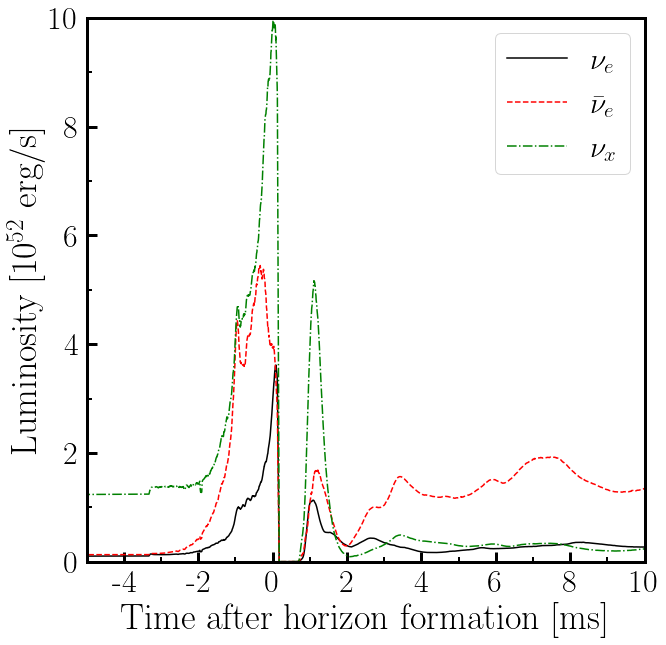}
\includegraphics[width=0.43\textwidth]{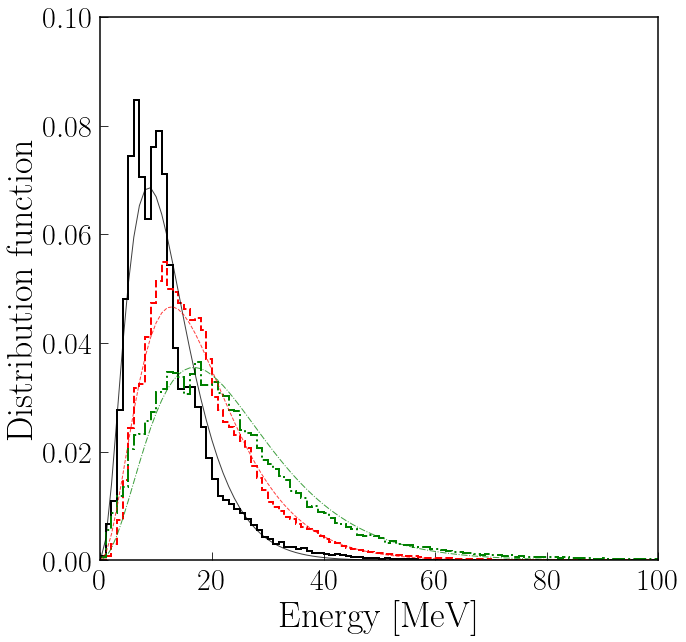}
 \caption{{\it Left}: Neutrino luminosity at the boundary of our computational domain in the LS220-178-106-vis simulation for $\nu_e$ (solid black), $\bar\nu_e$ (dashed red), and $\nu_x$ (all species combined, dot-dashed green). The drop at the merger time is due to an extension of the computational domain. We note that the neutrino luminosity remains low over the entire evolution, by the standard of neutron star mergers. The other simulations have comparable peaks before apparent horizon formation, and even lower luminosity at later times. {\it Right}: Energy spectrum of the escaping neutrinos, in $1\,{\rm MeV}$ bins. For neutrinos emitted close to the surface of the star (low-energy $\nu_e$), the structure of the bins used in the simulation remains very visible at the outer boundary. Each packet is weighted by the number of neutrinos that it represents (not by energy) and we show only the neutrinos leaving the grid after collapse of the remnant (including all neutrinos does not meaningfully change these results). We additionally plot for each species a Fermi-Dirac distribution with the same average energy as the observed neutrinos ($[12.3,18.1,23.8]$ MeV).}
\label{fig:NuLum}
\end{figure*}

Fig.~\ref{fig:NuLum} shows the neutrino luminosity as a function of time for system LS220-178-106. Before merger, we see steady emission of heavy lepton neutrinos ($\nu_x$) at $\sim 10^{52}\,{\rm ergs/s}$. That emission is largely due to heating of the neutron star surface by numerical viscosity.  Indeed, the difference in luminosity between simulations with and without viscous heating is O($1\%$). On the other hand, decreasing the grid spacing by $20\%$ leads to a comparable decrease in the luminosity (as this effect is due to the surface discontinuity at the surface of the neutron star, this slow first order convergence is not unexpected). Neutrino emission at this stage is dominated by pair processes in hot surface regions, and these processes are neglected for electron type neutrinos. Accordingly, emission of $\nu_e$ and $\bar\nu_e$ is an order of magnitude lower than emission of $\nu_x$.

About $2\,{\rm ms}$ before the detection of an apparent horizon, the luminosity quickly rises due to the increased temperature in the colliding neutron stars. There is nearly no time delay on this rise on Fig.~\ref{fig:NuLum}, because we measure the energy of neutrinos leaving the computational grid and, before collapse, our finite difference grid only covers the central region where the neutron stars are colliding. This increased neutrino emission is very brief: just $2\,{\rm ms}$ post-merger, all neutrino luminosities drop below $\sim 10^{52}\,{\rm ergs/s}$ (the gap in the neutrino luminosity on Fig.~\ref{fig:NuLum} is due to the extension of the grid after formation of a black hole).

At later times, an accretion disk forms around the remnant black hole, with fluid temperatures of a few MeV. That disk emits neutrinos with luminosity $\sim 2\times 10^{52}\,{\rm ergs/s}$ from $3\,{\rm ms}$ post-merger to the end of our simulation. During that time, the neutrino luminosity
is dominated by $\bar\nu_e$ emission, leading to the slow protonization of the post-merger accretion disk. The late time luminosity is only mildly impacted by viscosity: the simulation without viscosity sees a $\sim 20\%$ decrease in the $\bar\nu_e$ and $\nu_x$ luminosity, and a $\sim 5\%$ decrease in $\nu_e$
luminosity.

Our closest comparison point for the neutrino luminosity is probably the equal mass, rapidly collapsing system simulated in~\cite{Radice:2021jtw}. Our luminosities are about a factor of $2-4$ lower at peak, but follow qualitatively the same trend in the inspiral-merger phase (when comparing the inspiral emission, one should note that Fig 12 of~\cite{Radice:2021jtw} shows emission of $\nu_\mu$ only, while we show all four heavy lepton species together). As the simulation from~\cite{Radice:2021jtw} does not form a massive disk, their post-merger emission falls off very rapidly. Simulations evolving massive disks forming around black holes do however find a rise in the neutrino luminosity to $\sim 10^{53-54}\,{\rm erg/s}$ over $\sim 10\,{\rm ms}$ time scales, followed by a decay on slightly longer timescales (see e.g.~\cite{Hayashi:2021oxy}). The post-merger remnant in this system is relativitely cold and not particularly massive. Its total neutrino luminosity is nearly an order of magnitude lower than that of~\cite{Hayashi:2021oxy}.

\begin{figure}
\includegraphics[width=0.9\columnwidth]{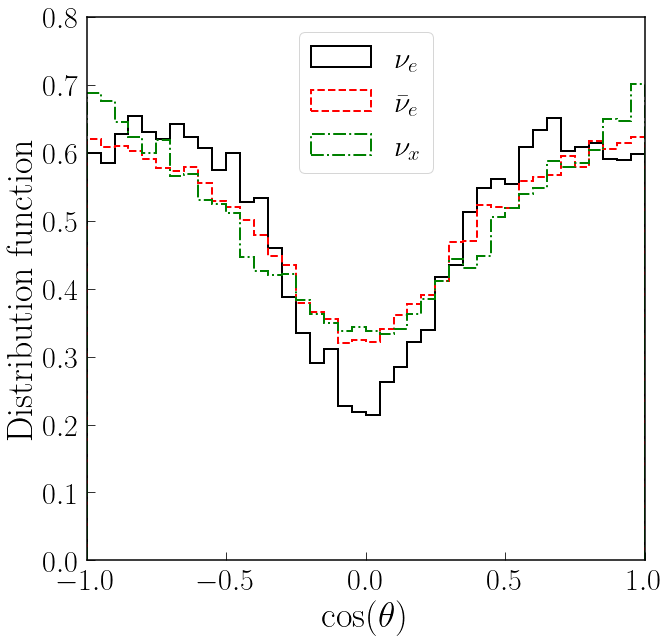}
 \caption{Distribution of neutrinos as a function of $\cos{\theta}$, with $\theta$ the angle with respect to the axis of rotation of the remnant.}
\label{fig:MuDist}
\end{figure}

Fig.~\ref{fig:NuLum} also shows the spectrum of neutrinos leaving the computational domain. After collapse to a black hole, we find the usual ordering of energies, with the average energy of $\nu_e$ being the lowest ($12.2\,{\rm MeV}$), then the energy of $\bar \nu_e$ ($18.1\,{\rm MeV}$), and finally the energy of $\nu_x$ ($23.8\,{\rm MeV}$). By comparing the neutrino spectra to what we would expect from a thermal distribution at a temperature chosen to match the average neutrino energy observed in our simulation, we see that the neutrino emission is nearly consistent with black-body emission (at $T_{\nu_e}=(3.9, 5.7,7.6)\,{\rm MeV}$ for $(\nu_e,\bar\nu_e,\nu_x)$). Fig.~\ref{fig:MuDist} shows the angular distribution of neutrinos leaving the remnant. We find a relatively isotropic distribution of neutrinos, with a mild preference for emission along the rotation axis of the remnant.

Given the low neutrino luminosity observed in our simulations, it is relatively unsurprising that the composition of our outflows remain extremely neutron rich. Whether the difference in the evolution of $Y_e$ between our results and~\cite{Bernuzzi:2020txg} is due to differences in the implementation of neutrino transport, differences in the chosen viscosity model, or differences in the numerical methods used is however difficult to assess with the limited number of simulations available, as these questions are very tightly intertwined (e.g. even if we knew that the neutrino luminosity were higher in~\cite{Bernuzzi:2020txg}, this could be due to the transport algorithm, to more viscous heating from the subgrid model, to a higher numerical viscosity, or to different treatments of shocks,...).

\subsection{Lagrangian Tracers}

\begin{figure}
\includegraphics[width=0.9\columnwidth]{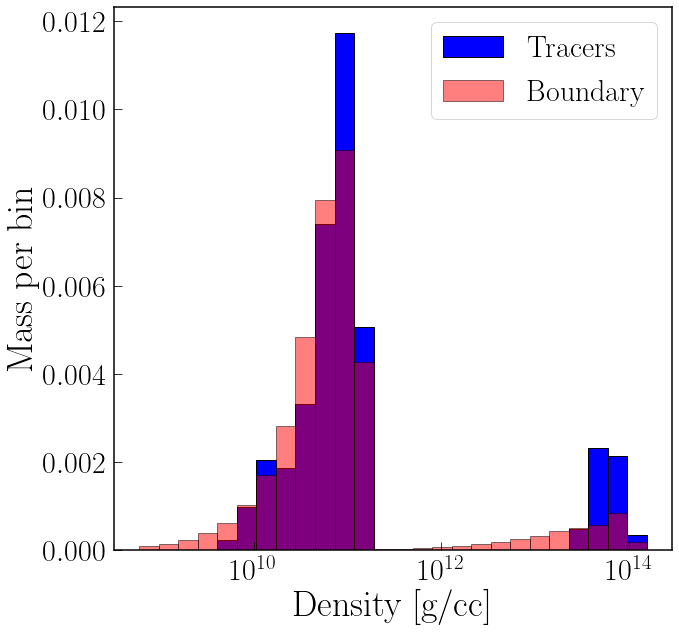}
\includegraphics[width=0.9\columnwidth]{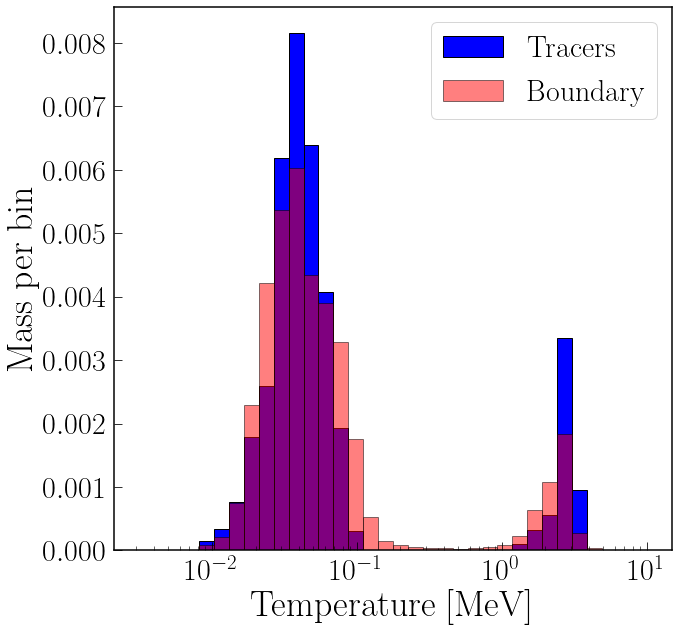}
 \caption{Density ({\it Top}) and temperature ({\it Bottom}) of the matter leaving the computational grid according to Lagrangian tracers following the fluid from the beginning of the inspiral (blue) or from observing directly matter fluxes on the grid boundary. Despite the long evolution of the tracers, the two remain in decent qualitative agreement, with the tracers slightly oversampling higher density regions.}
\label{fig:Tracers}
\end{figure}

As discussed in Sec.~\ref{sec:tracers}, simulation LS220-178-106 was performed with Lagrangian tracer particles. Each particle in the simulation represents a baryon mass of $10^{-5}M_\odot$. These tracers are initialized by sampling the neutron star matter at the beginning of the simulation, about $20\,{\rm ms}$ before merger. The main use of these tracers is to provide a history of the density, temperature, and composition of the matter leaving the grid. Indeed, these quantities are needed to use the result of numerical simulations as input for nuclear reaction networks, and thus to predict the outcome of nucleosynthesis in a given merger. Considering the relatively long evolution of these tracers, and the fact that their evolution in time is not guaranteed to be self-consistent with the evolution of the fluid, it is worth studying whether the tracers leaving the computational grid still provide a good sampling of the outflows. With that in mind, we compare the properties of the tracers leaving the computational grid with a direct measurement of the properties of matter outflows on the boundary of our computational domain. 

First, we note that there is good agreement for the total amount of matter leaving the computational domain. We measure $0.0374M_\odot$ leaving the grid at the boundary, and a total of $3795$ tracers leaving the grid (ignoring both matter and tracers falling into the black hole). Estimating the expected statistical uncertainty on the number of tracers leaving the grid to be $\sigma \approx \sqrt{3797}=62$, the tracers predict that $(0.0380\pm 0.0006)M_\odot$ of material leaves the grid, and the difference between the two methods is thus not significant. A more detailed analysis of the properties of the outflows shows some slightly more significant differences. On Fig.~\ref{fig:Tracers}, we plot the distribution of density and temperature for the matter leaving the grid, according to both the tracer particles and measurements at the boundary. There is fairly reasonable qualitative agreement between the two methods, but with differences that are now significantly outside of our (roughly) estimated statistical errors. The two peaks in the distribution correspond to matter ejected during merger (higher density and temperature) and after collapse to a black hole (lower density and temperature); they are in large part due to the fact that the matter outflows are observed closer to the remnant during merger. In both regimes, however, it appears that the tracers are slightly oversampling the densest regions of the outflows. A similar graph for the electron fraction $Y_e$, on the other hand, shows no noticeable difference between the two methods (but the distribution of $Y_e$ in these simulations is extremely narrow). In these binary neutron star mergers, it thus appears reasonable to use tracer particles evolved as described in Sec.~\ref{sec:tracers} as input for nuclear reaction networks, considering the significantly larger sources of uncertainty coming from unknown nuclear physics and from limited physics in the simulations themselves. We note already that this is not guaranteed to be true for all systems. We have recently performed a simulation of a very asymmetric black hole-neutron star binary, and in that system tracers initialized before disruption did {\it not} provide a good sampling of the dynamical ejecta. We also performed a separate simulation of LS220-178-106 in which the tracers were initalized at the time of contact between the two neutron stars. Despite a later initialization of the tracers, we find worse agreement between the tracers and matter outflows measured at the boundary ($\sim 5\%$ error in the outflow mass, or about three times the expected statistical error). We thus note the importance of obtaining some cross-checks for the validity of the sampling performed by tracer particles before using them for e.g. nucleosynthesis calculations. 

\begin{figure*}
\includegraphics[width=0.9\textwidth]{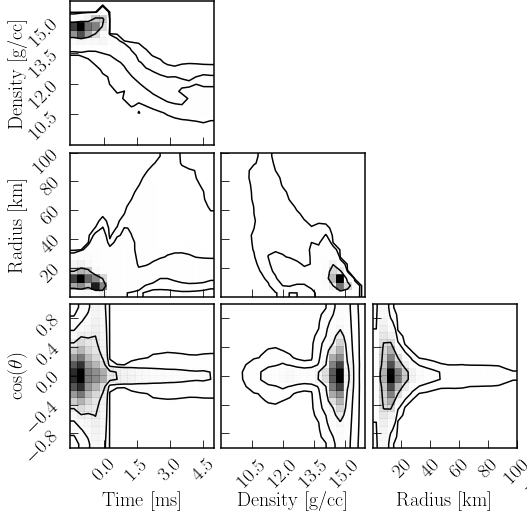}
 \caption{Properties of the fluid according to the Lagrangian tracers. Here, $\theta$ is the angle with respect to the rotation axis of the remnant. We clearly see the increase in density of the core before collapse to a black hole (at $t=0$ in these coordinates), followed by the formation of an accretion disk at radius $(20-30)\,{\rm km}$ at density $\sim 10^{12}\,{\rm g/cm^3}$. As before, we show $(1,2,3)-\sigma$ contours of the various distributions, weighted by mass. We note that tracers that cross the apparent horizon of the black hole are removed from the simulation.}
\label{fig:AllTracers}
\end{figure*}

The tracers can also be used to study the evolution of the merger remnant itself. In Fig.~\ref{fig:AllTracers}, we show the distribution of tracers as a function of time, radius, polar angle, and fluid density. Most of the tracers naturally disappear into the black hole within $\sim 0.1\,{\rm ms}$ post-collapse, but the remaining tracers allow us to follow the evolution of the properties of  the forming accretion disk.

\section{Conclusions}

We perform simulations of four distinct high mass binary neutron star systems using different mass ratios and equations of state. All of the systems studied here immediately collapse to a black hole upon merger. As expected, the more symmetric systems do not result in any mass ejection, and do not lead to the formation of massive accretion disks. More asymmetric systems, on the other hand, eject up to $\sim 0.01M_\odot$ in a cold tidal tail, with even more matter in a bound disk and tidal tail (up to $\sim 0.15M_\odot$). Our simulations include state-of-the art general relativistic radiation hydrodynamics. We neglect magnetic fields, but do include in some simulations a subgrid viscosity model for the Kelvin-Helmotz instability (during merger) and the MRI (post-merger). The former has a negligible impact on the outcome of our simulations.

Our most asymmetric systems involve the merger of a $1.06M_\odot$ neutron star with its $1.78M_\odot$ companion. These simulations, performed with both the LS220 and SFHo equations of state exactly match systems previously studied by Bernuzzi {\it et al}~\cite{Bernuzzi:2020txg} with different methods for the evolution of the metric, fluid, neutrinos, and a different subgrid viscosisy model. Despite these significant differences, we find excellent agreement in the mass and spin of the black hole remnant, the bound mass, and the ejected mass (for the system with significant ejecta). The main noticeable differences are a factor of $2$ in the ejected mass for the system producing only $(0.001-0.002)M_\odot$ of dynamical ejecta, and a lower electron fraction in our outflows for simulations including viscosity. Considering the many differences between our numerical codes and in the microphysics included in the simulations, this is a very reassuring result for the robustness of our predictions for the cold tidal ejecta from binary neutron star mergers.

Besides this important verifications of the consistency of two distinct numerical relativity code, our simulations provide a range of observables of use for the modeling of neutron star mergers and of their electromagnetic properties. Namely, we have detailed sampling of the matter ejected by each binary merger and of the neutrinos leaving the computational grid, tracer particles providing the history of $\sim 10^5$ fluid elements in the simulation, as well as snapshots of the final post-merger remnant available for long post-merger evolution of these systems.

For the binary parameters considered here, the only significant source of outflows over the timescale of our evolutions (up to $10\,{\rm ms}$ post black hole formation) is the cold tidal ejecta. We notice only negligeable amounts of hot outflows and of post-merger disk winds. We note however that our simulations would not capture the production of magnetically-driven winds from the post-merger remnant that are expected to be ejected during the $\sim 0.1\,{\rm s}$ following the merger. Our viscosity model would only capture the viscous outflows observed on seconds timescales, i.e. over much longer timescale than the simulations presented in this manuscript.

Neutrinos do not appear to play a particularly important role in the properties of merger outflows in these systems, as expected for cold dynamical ejecta. Neutrinos do however impact the composition of the post-merger remnant. Additionally, our simulations allowed us to verify that cheap evolutions of black hole-disk system with Monte-Carlo radiation transport are indeed possible. In a system like this one, which is largely devoid of hot, dense, optically thick regions, Monte-Carlo radiation transport is significantly cheaper than the use of e.g. our two-moment scheme. As the main issues encountered by Monte-Carlo schemes are also in those dense regions, the Monte-Carlo scheme is also likely to be significantly more accurate than a two-moment scheme, and provides us with information about e.g. the spectrum of neutrinos.

Our LS220 simulation includes evolution of Lagrangian tracer particles, made available to provide the history of the matter leaving the computational domain (as well as matter that remains on the grid if necessary). In this manuscript, we focus on the accuracy of the evolution of these tracers, and show that tracers do accurately capture the mass ejected by the system (to within $\sim 2\%-5\%$ here, depending on the simulation). Slightly larger errors are observed for more detailed fluid properties, though those remain small compared to the many other sources of uncertainty affecting the properties of the outflows (e.g. for LS220-178-106 $\sim 10\%$ for the viscous model and $\sim 25\%$ for the impact of r-process heating).

Finally, we take advantage of these simulations to compare different methods available to estimate the amount of mass ejected by the merger, making different assumptions about the conversion of thermal and nuclear energy into kinetic energy in the outflows. We confirm that the choices made in the definition of the outflows have a major impact on the amount of matter ejected by our simulations. By comparing our two best estimates of the ejected mass, we estimate uncertainties of $(10-50)\%$ in the predicted ejected mass from this effect alone.

\begin{acknowledgments}
F.F. gratefully acknowledges support from the Department of Energy, Office of Science, Office of Nuclear Physics, under contract number
DE-AC02-05CH11231 and from the NSF through grant AST-2107932. M.D. gratefully acknowledges support from the NSF through grant PHY-2110287.  M.D. and F.F. gratefully acknowledge support from NASA through grant 80NSSC22K0719. M.S. acknowledges funding from the Sherman Fairchild Foundation
and by NSF Grants No. PHY-1708212, No. PHY-1708213, and No. OAC-1931266
at Caltech.  L.K. acknowledges funding from the Sherman Fairchild Foundation
and by NSF Grants No. PHY-1912081, No. PHY-2207342, and No. OAC-1931280
at Cornell. Computations for this manuscript were performed on the Plasma cluster, a Cray CS500 supercomputer at UNH supported by the NSF MRI program under grant AGS-1919310, and on the Wheeler cluster at Caltech, supported by the Sherman Fairchild Foundation. The authors acknowledge the Texas Advanced Computing Center (TACC) at The University of Texas at Austin and the NSF for providing resources on the Frontera cluster~\cite{10.1145/3311790.3396656} that have contributed to the research results reported within this paper.

\end{acknowledgments}

\bibliography{References/References.bib}

\end{document}